\documentclass[twocolumn]{aastex63}
\usepackage{amsmath}

\hypersetup{linkcolor=magenta,citecolor=blue,filecolor=cyan,urlcolor=purple}

\received{January 21, 2021} 
\revised{March 22, 2021} 
\accepted{April 16, 2021} 
\submitjournal{ApJ}

\shorttitle{CME from an Extended Filament Channel}
\shortauthors{Lynch et al.}

\begin{document}

\title{Modeling a Coronal Mass Ejection from an Extended Filament Channel. \\ I. Eruption and Early Evolution}

\correspondingauthor{Benjamin~J.~Lynch}
\email{blynch@berkeley.edu}

\author[0000-0001-6886-855X]{Benjamin~J.~Lynch}
\affil{Space Sciences Laboratory, University of California--Berkeley,
        Berkeley, CA 94720, USA}

\author[0000-0001-6590-3479]{Erika~Palmerio}
\affil{Space Sciences Laboratory, University of California--Berkeley,
        Berkeley, CA 94720, USA}
\affil{CPAESS, University Corporation for Atmospheric Research, Boulder, CO 80301, USA}
\affil{Department of Physics, University of Helsinki, FI-00014 Helsinki, Finland}

\author[0000-0002-4668-591X]{C.~Richard~DeVore}
\affil{Heliophysics Science Division, NASA Goddard Space Flight Center, Greenbelt, MD 20771, USA}

\author[0000-0001-8975-7605]{Maria~D.~Kazachenko}
\affil{Department of Astrophysical and Planetary Sciences, University of Colorado--Boulder, Boulder, CO 80305, USA}
\affil{National Solar Observatory, University of Colorado--Boulder, Boulder, CO 80303, USA}

\author[0000-0002-9493-4730]{Joel~T.~Dahlin}
\affil{Heliophysics Science Division, NASA Goddard Space Flight Center, Greenbelt, MD 20771, USA}

\author[0000-0003-1175-7124]{Jens~Pomoell}
\affil{Department of Physics, University of Helsinki, FI-00014 Helsinki, Finland}

\author[0000-0002-4489-8073]{Emilia~K.~J.~Kilpua}
\affil{Department of Physics, University of Helsinki, FI-00014 Helsinki, Finland}

\begin{abstract}

We present observations and modeling of the magnetic field configuration, morphology, and dynamics of a large-scale, high-latitude filament eruption observed by the Solar Dynamics Observatory. We analyze the 2015 July 9--10 filament eruption and the evolution of the resulting coronal mass ejection (CME) through the solar corona. The slow streamer-blowout CME leaves behind an elongated post-eruption arcade above the extended polarity inversion line that is only poorly visible in extreme ultraviolet (EUV) disk observations and does not resemble a typical bright flare-loop system. 
Magnetohydrodynamic (MHD) simulation results from our data-inspired modeling of this eruption compare favorably with the EUV and white-light coronagraph observations.
We estimate the reconnection flux from the simulation's flare-arcade growth and examine the magnetic-field orientation and evolution of the erupting prominence, highlighting the transition from an erupting sheared-arcade filament channel into a streamer-blowout flux-rope CME.
Our results represent the first numerical modeling of a global-scale filament eruption where multiple ambiguous and complex observational signatures in EUV and white light can be fully understood and explained with the MHD simulation.
{In this context, our findings also suggest that the so-called ``stealth CME” classification, as a driver of unexpected or ``problem'' geomagnetic storms, belongs more to a continuum of observable/non-observable signatures than to separate or distinct eruption processes.}

\end{abstract}




\section{Introduction} \label{sec:intro}

{Large-scale filament eruptions that drive coronal mass ejections (CMEs) are some of the most spectacular energetic and dynamic transients in the solar corona. A well-known property of CME source regions is that the magnetic free energy required to power a solar eruption is concentrated above radial magnetic field polarity inversion lines (PILs) in the form of stressed, sheared, and/or twisted magnetic field structures \citep[e.g., see][and references therein]{Patsourakos2020}.
\citet{Pevtsov2012} examined several large-scale filament channels, i.e., long PILs in the underlying photospheric magnetic field distribution with highly-sheared coronal magnetic fields, including those without any discernible filament or prominence material, and showed that these types of coronal structures were responsible for slow- to moderate-speed CME events.
Typical low-coronal eruption signatures (e.g. filament eruptions in H$\alpha$ or extreme ultra-violet (EUV) and  ultra-violet (UV) flare ribbons and post-eruption arcades, soft X-ray sigmoid-to-arcade transitions, and/or large-scale coronal dimmings), when followed by halo or partial-halo CMEs, can act as a warning for potential geomagnetic storms from Earth-impacting CMEs anywhere from 2--5 days in advance.}

{The concept of ``stealth CMEs'' was first introduced by \citet{Robbrecht2009}, who used the multi-spacecraft viewing perspective of the Solar Terrestrial Relations Observatory (STEREO) spacecraft to analyze a classic, slow streamer-blowout eruption observed at the eastern limb in white-light coronagraph observations from STEREO-A (STA) and with a head-on view from STEREO-B (STB). The event had none of the corresponding on-disk signatures that are usually associated with CMEs. 
The ``stealth CME'' classification is necessarily somewhat subjective, based as it is on the interpretation of remote-sensing data and its quality and processing methods. Consequently, we suggest that using the adjective \emph{stealthy} to describe any individual CME event that appears to be missing one or more of the expected on-disk, low-coronal eruption signatures would be a more appropriate nomenclature. 
From a space-weather perspective, stealth CME impacts are particularly difficult to forecast as potentially geoeffective ICMEs precisely because they lack the usual on-disk eruption signatures \citep[e.g.][]{Nitta2017}.
The resulting ``problem'' geomagnetic storms are often associated with unexpected ICME interactions with Earth's magnetosphere.}

Our previous simulation \citep{Lynch2016b} of the initiation and low-coronal evolution of a slow, streamer-blowout CME based on the \citet{Robbrecht2009} stealth CME event showed excellent qualitative agreement with the STEREO-A coronagraph observations. The magnetohydrodynamic (MHD) model successfully replicated several features, including the height and morphology of the X-point flare current sheet in the STA/COR1 field of view (STA saw the stealth CME above the eastern limb), the three-part structure of the flux-rope CME in synthetic running-difference images, and the height--time and velocity profiles of the stealth CME propagation through the STA/COR2 field of view (${\sim} 15\,R_\odot$). Additionally, the simulation's eruptive flare reconnection gradually released ${\sim}10^{30}$~erg of magnetic energy over ${\gtrsim}20$~hr and over such a large spatial extent that the estimated energy flux into the post-eruption arcade system was unlikely to cause observable temperature increase or emission enhancement, providing a natural explanation for the lack of ``flare-like'' low-coronal signatures. On this basis, we argued that the initiation mechanism for stealth CMEs is not fundamentally different from most slow streamer-blowout CMEs: they simply represent the lowest-energy range of the CME distribution. 

In this paper, we extend this argument and build upon the \citet{Lynch2016b} simulation results by modeling the 2015 July 9--10 slow streamer-blowout eruption that originated from an extended, high-latitude filament channel that appeared to span most of the solar disk. There are a number of observable on-disk and off-limb low-coronal signatures associated with this eruption, so this event {cannot be considered a stealth CME akin to the} \citet{Robbrecht2009} case. However, as described below, our event's low coronal signatures are somewhat ambiguous when taken individually and present a qualitatively weaker indication that a possibly geoeffective, Earth-directed eruption has taken place, especially compared to most flare-associated CMEs from active regions. Our numerical modeling shows that this particular set of low-coronal signatures can be understood as weak or {quasi-stealthy} manifestations of the expected flare ribbons and post-eruption flare arcade resulting from a CME eruption following the \citet{Lynch2016b} scenario. 

The paper is organized as follows. In Section~\ref{sec:obs} we describe the 2015 July 9--10 CME event and present the timeline of the non-standard low-coronal signatures of the filament eruption ($\S$\ref{subsec:disk_obs}) and its morphology and evolution in coronagraph observations ($\S$\ref{subsec:corona_obs}). In Section~\ref{sec:simsetup} we describe the MHD numerical modeling methodology ($\S$\ref{subsec:arms}), the initial potential-field source-surface configuration derived from the magnetogram synoptic map for Carrington Rotation (CR) 2165 and background solar wind ($\S$\ref{subsec:init}), and the boundary flows used to energize the model filament-channel configuration ($\S$\ref{subsec:flows}). In Section~\ref{sec:simresults} we present the simulation results, examining the filament-channel magnetic-field structure and its evolution during CME initiation ($\S$\ref{subsec:filament}), the eruption-related dimming signatures and development of the stealthy flare ribbons and post-eruption arcade system ($\S$\ref{subsec:arcade}), and the synthetic white-light coronagraph morphology of the slow eruption ($\S$\ref{subsec:coronagraph}), with direct comparisons to the observations. {Finally, in Section~\ref{sec:disc} we summarize and discuss the implications of our results for the observed dynamics of CME eruptions in the solar corona and space weather forecasting.} 


\begin{figure*}[t]
	\centering{
	\includegraphics[width=0.95\textwidth]{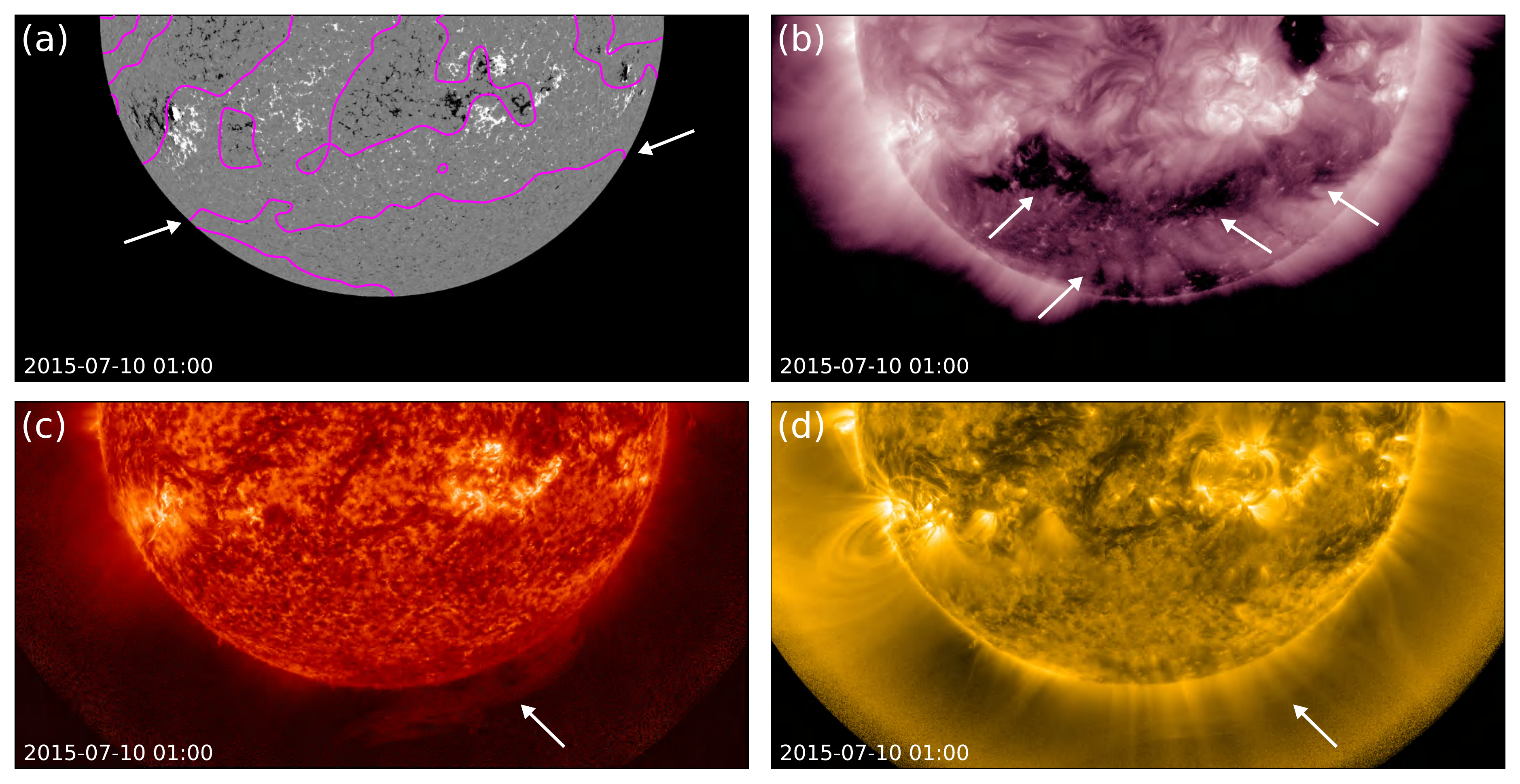}
	}
	\caption{Overview of the 2015 July 9--10 filament eruption in different SDO data sets. (a): HMI line-of-sight (LOS) magnetogram overlaid with the global PILs (in magenta). The PIL involved in the eruption under study is marked with arrows.
	(b): AIA 211~{\AA} image showing {multiple eruption-related coronal dimming areas (marked with arrows)}.
	(c): AIA 304~{\AA} image showing the erupting filament material off limb (marked with an arrow).
	(d): AIA 171~{\AA} image showing opening loops off limb (marked with an arrow).
	The animated version of this figure runs from 2015 July 9 at 12:00~UT to 2015 July 10 at 21:00~UT and shows the whole eruption process, from the activation of the high-latitude filament to the full development of the post-eruptive signatures.\\
	(An animation of this figure is available.)}
	\label{f1}
\end{figure*}

\section{Overview of the 2015 July 9--10 Eruption} \label{sec:obs}

\subsection{Solar Disk Observations} \label{subsec:disk_obs}

The CME that we analyze originated from the quiet Sun in the southern hemisphere in July 2015. The CME source region was an extended, high-latitude filament channel that spanned the whole Earth-facing disk in longitude. The large-scale, sheared-arcade field above the global PIL is a common feature of extended filament channels on the Sun \citep{MacKay2010,Pevtsov2012}. The eruption followed the standard scenario of every large prominence eruption, albeit very slowly: the stressed field of an energized filament channel slowly rose and eventually transitioned to runaway expansion that drove reconnection beneath the erupting structure \citep[e.g.,][]{Sterling2003,Su2012,Parenti2014,Su2015}.

Figure~\ref{f1} shows a still frame of the filament eruption on 2015 July 10 at 01:00 UT, using four different data sets from the Solar Dynamics Observatory \citep[SDO;][]{Pesnell2012}: magnetograph data from the Helioseismic and Magnetic Imager \citep[HMI;][]{Scherrer2012} and EUV data in the channels at 211, 304, and 171~{\AA} from the Atmospheric Imaging Assembly \citep[AIA;][]{Lemen2012}. The animated version of Figure~\ref{f1} showcases on-disk observations of the eruption over the course of 33~hours (from 2015 July 9 at 12:00~UT to 2015 July 10 at 21:00~UT). The high-latitude PIL associated with the eruption is visible in Figure~\ref{f1}(a) extending from limb to limb. Magnetogram data show that the region of quiet Sun to the north of the PIL is predominantly of positive polarity, while the southern region (in the vicinity of the polar coronal hole) is predominantly negative. Each of the AIA channels displayed in Figure~\ref{f1}(b)-(d) shows different manifestations of the gradual filament eruption and post-eruptive evolution. 

Before eruption onset, filament material can be identified in both the 211~{\AA} and 304~{\AA} channels. This material is seen to activate and slowly rise starting at around 16:00~UT on July 9. A couple of hours later, at about 18:30~UT, 171~{\AA} imagery (Figure~\ref{f1}(d)) reveals a coronal loop opening off limb to the east of the PIL. This is likely the trace of the eastern leg of the extended filament (anchored behind the eastern limb) lifting off from the Sun. The western spine of the filament, on the other hand, begins to be visible in the 304~{\AA} channel (Figure~\ref{f1}(c)) off limb to the southwest of the solar disk starting around 22:00~UT, with its corresponding opening loops also visible in 171~{\AA} data (Figure~\ref{f1}(d)). Clear coronal dimmings \citep[signatures of CME mass leaving the solar corona; see e.g.][]{Thompson2000} can be seen to form in 211~{\AA} imagery (Figure~\ref{f1}(b)) around 00:00~UT on July~10 and progressively migrate away from the PIL. Furthermore, an east-to-west on-disk signature reminiscent of flare-ribbon and post-eruption-arcade evolution is noticeable in the 171~{\AA} channel (some features are visible at 304~{\AA} as well), starting at around 04:00~UT and persisting well beyond July~10 at 21:00~UT.

This extremely large-scale, gradual filament eruption does not occur simultaneously along the entire PIL, rather it erupts asymmetrically \citep{Tripathi2006,LiuR2009,McCauley2015}. The animated version of Figure~\ref{f1} shows the eruption signatures starting at the eastern limb, progressing across the disk face, and finishing at the west limb without the impulsive, explosive character of strong-field active-region CMEs.

\begin{figure*}[tbh]
    \centering{
    \includegraphics[width=0.95\textwidth]{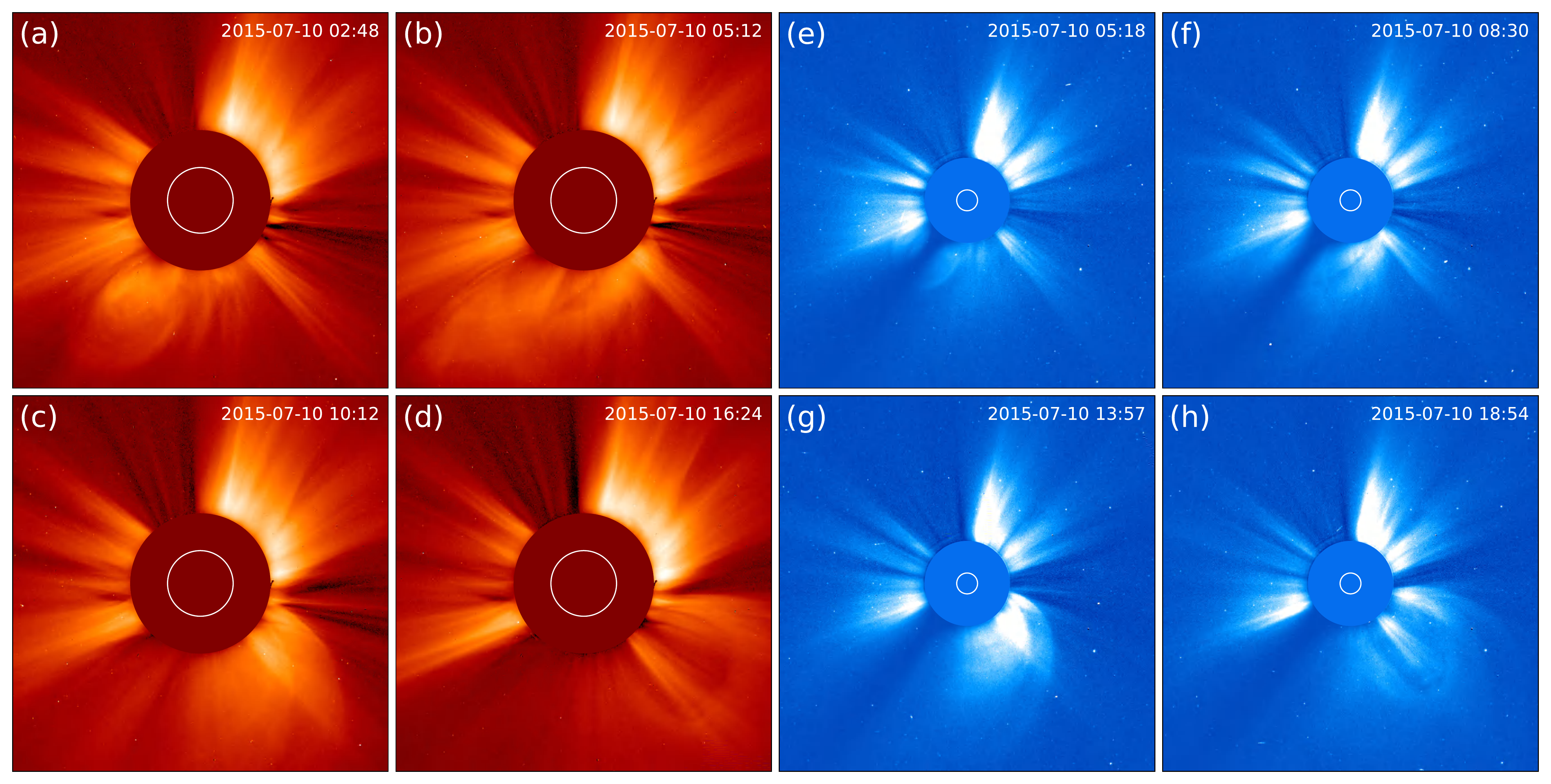}
    }
	\caption{Overview of the 2015 July 9--10 CME eruption as seen in the LASCO coronagraphs. (a)--(d): LASCO/C2 data; (e)--(h): LASCO/C3 data. The animated version of this figure runs from 2015 July 9 at 16:00~UT to 2015 July 11 at 00:00~UT and shows the whole passage of the CME through the fields of view of both coronagraphs.\\
	(An animation of this figure is available.)  }
	\label{f2}
\end{figure*}

\subsection{Coronagraph Observations} \label{subsec:corona_obs}

After its eruption from the Sun, the 2015 July 9--10 CME appeared in coronagraph imagery from the Large Angle Spectroscopic Coronagraph \citep[LASCO;][]{Brueckner1995} onboard the Solar and Heliospheric Observatory \citep[SOHO;][]{Domingo1995}. The east-to-west asymmetry of the eruption at the Sun is well reflected through the solar corona, where the CME proceeded slowly over the course of about one day. Figure~\ref{f2} shows an overview of the passage of the CME through the C2 and C3 coronagraphs in the LASCO package. The animated version of Figure~\ref{f2} showcases the evolution of the CME through the solar corona over the course of 32~hours (from 2015 July 9 at 16:00 UT to 2015 July 11 at 00:00 UT). 

In the C2 imagery shown in the animated version of Figure~\ref{f2}, the streamer to the southeast of the solar disk swells until the appearance of the leading edge of the CME around 20:00~UT on July~9. This portion of the eruption likely corresponds to the activation of the eastern leg of the filament (described in $\S$\ref{subsec:disk_obs}) and appears in coronagraph data as a slow streamer blowout that is quite narrow (Figure~\ref{f2}(a) and (e)). This ejected material to the southeast of the disk later (at around 03:00 UT on July 10) is followed by the appearance of a more extended, asymmetric feature that sweeps a large part of the southern corona from east to west (Figure~\ref{f2}(b,c) and (f,g)). This portion of the eruption likely corresponds to the liftoff of the western filament spine (described in $\S$\ref{subsec:disk_obs}). Finally, at around 14:30~UT on July 10, a structure reminiscent of a three-part-CME cavity appears to the southwest of the solar disk and propagates outwards with the western leg of the CME body (Figure~\ref{f2}(d) and (h)). This feature does not appear to have a corresponding counterpart in solar-disk imagery; nevertheless, the timing and loop-like morphology suggest that it may be associated with the trailing edge of the density-depleted, erupting flux-rope cavity.

In brief, the 2015 July 9--10 CME appeared in coronagraph data as a long-duration (${\sim}24$~hr) event that was composed of three portions: a narrow streamer blowout to the southeast; an asymmetric, large ejection throughout the southern hemisphere; and a flux-rope cavity to the southwest.


\section{Modeling the Pre-Eruption Corona} \label{sec:simsetup}

\subsection{Numerical Methods} \label{subsec:arms}

{Our numerical simulation is run with the Adaptively Refined MHD Solver \citep[ARMS;][]{DeVore2008} code. ARMS solves the 3D nonlinear, time-dependent equations of ideal MHD. It is based on a finite-volume, flux-corrected transport algorithm \citep{DeVore1991} that advances the equations for the conservation of mass, momentum, and energy, as well as the evolution of the magnetic field and electric currents throughout the system. ARMS utilizes the adaptive-mesh toolkit PARAMESH \citep{MacNeice2000} to enable efficient multiprocessor parallelization and dynamic, solution-adaptive computational block refinement.}

In this simulation, we solve the equations of ideal MHD in an isothermal atmosphere. This formulation does not employ an explicit physical resistivity term in the induction equation, but we note that there are stabilizing numerical-diffusion terms that introduce an effective resistivity at the computational grid scale. The numerical diffusion facilitates magnetic reconnection in regions where current-sheet features and their associated strong gradients have been compressed to the grid scale.

The spherical computational domain uses logarithmic grid spacing in $r$ and uniform grid spacing in ($\theta, \phi$). The domain extends over $r \in \left[1\,R_\odot, 30\,R_\odot\right]$, $\theta \in \left[ 11.25^{\circ}, 168.75^{\circ} \right]$ ($\pm 78.75^{\circ}$ in latitude), and $\phi \in \left[-180^{\circ}, +180^{\circ} \right]$ (longitude).  The initial grid consists of $7 \times 7 \times 15$ blocks with 8$^3$ grid cells per block. Three additional levels of static grid refinement are allowed.
The level-3 refinement extends over $r \in \left[1\,R_\odot, 6.984\,R_\odot\right]$ for all ($\theta, \phi$), and a spherical wedge of level-4 refinement centered on the high-latitude filament channel extends over $r \in \left[1\,R_\odot, 2.642\,R_\odot\right]$, $\theta \in \left[82.65^{\circ}, 168.75^{\circ}\right]$, and $\phi \in \left[ 0^{\circ}, 126^{\circ}\right]$.
This region corresponds to an effective maximum resolution of $448 \times 448 \times 960$. The level-4 grid cells have angular extent $0.352^{\circ} \times 0.375^{\circ}$ in $\theta, \phi$ and radial extent of $\Delta r = 0.00762\,R_\odot$ at the lower boundary.

The boundary conditions are as follows. The $\phi$ boundary is periodic. The $\theta$ boundaries are reflecting for the normal component, and free-slip for the tangential components (i.e.\ zero-gradient between the boundary interior cell and the guard cells). On the lower $r$ boundary, the magnetic field is line-tied with tangential velocities set to zero. The normal velocity can be positive, but not negative, thus allowing the radial guard cells to provide a positive mass flux into the computational domain. The outer $r$ boundary is flow-through (zero-gradient) for the normal velocity component, and ``half-slip'' for the tangential components (i.e.\ the tangential components are set to zero in the guard cells).

\subsection{Global Coronal Magnetic Field and Solar Wind} \label{subsec:init}

\begin{figure*}[!t]
    \centering{
	\includegraphics[width=0.95\textwidth]{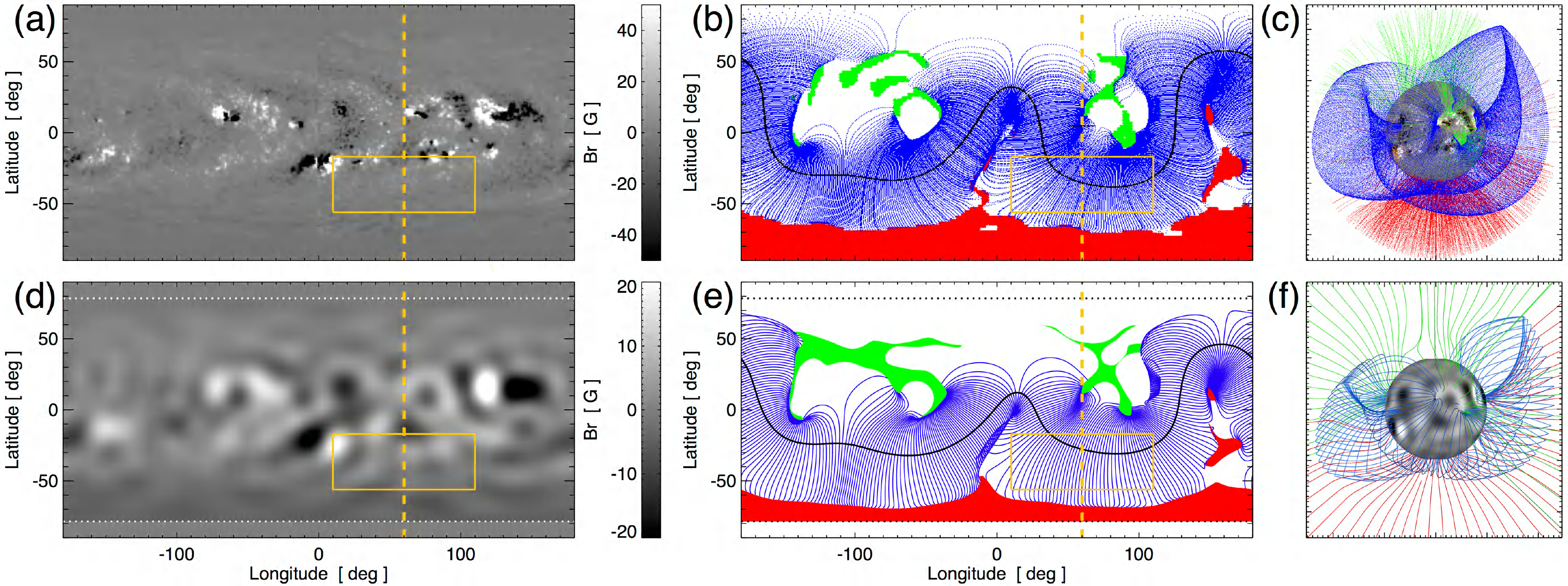}
	}
	\caption{(a): NSO/GONG synoptic map for CR~2165. The day of observation (2015 July 10) is shown as the yellow vertical dashed line and the high-latitude PIL source region for the prominence eruption is the yellow rectangle. (b): NSO/GONG PFSS visualization of coronal hole regions and helmet streamer configuration. Green (red) indicate positive (negative) polarity of the radial field and the black line shows the location of the HCS at the $2.5\,R_\odot$ source surface. (c): NSO/GONG PFSS Earth-view visualization on 2015 July 10. (d): ARMS $B_r$ distribution at the lower boundary. The dashed horizontal lines indicate the latitudinal boundaries of the computational domain. (e): ARMS coronal-hole and helmet-streamer configuration at $t=0$~hr. (f): Earth-view visualization of the ARMS initial PFSS field structure.
	}
	\label{f3}
\end{figure*}
\begin{figure*}[t!]
	\centering{ \includegraphics[width=0.95\textwidth]{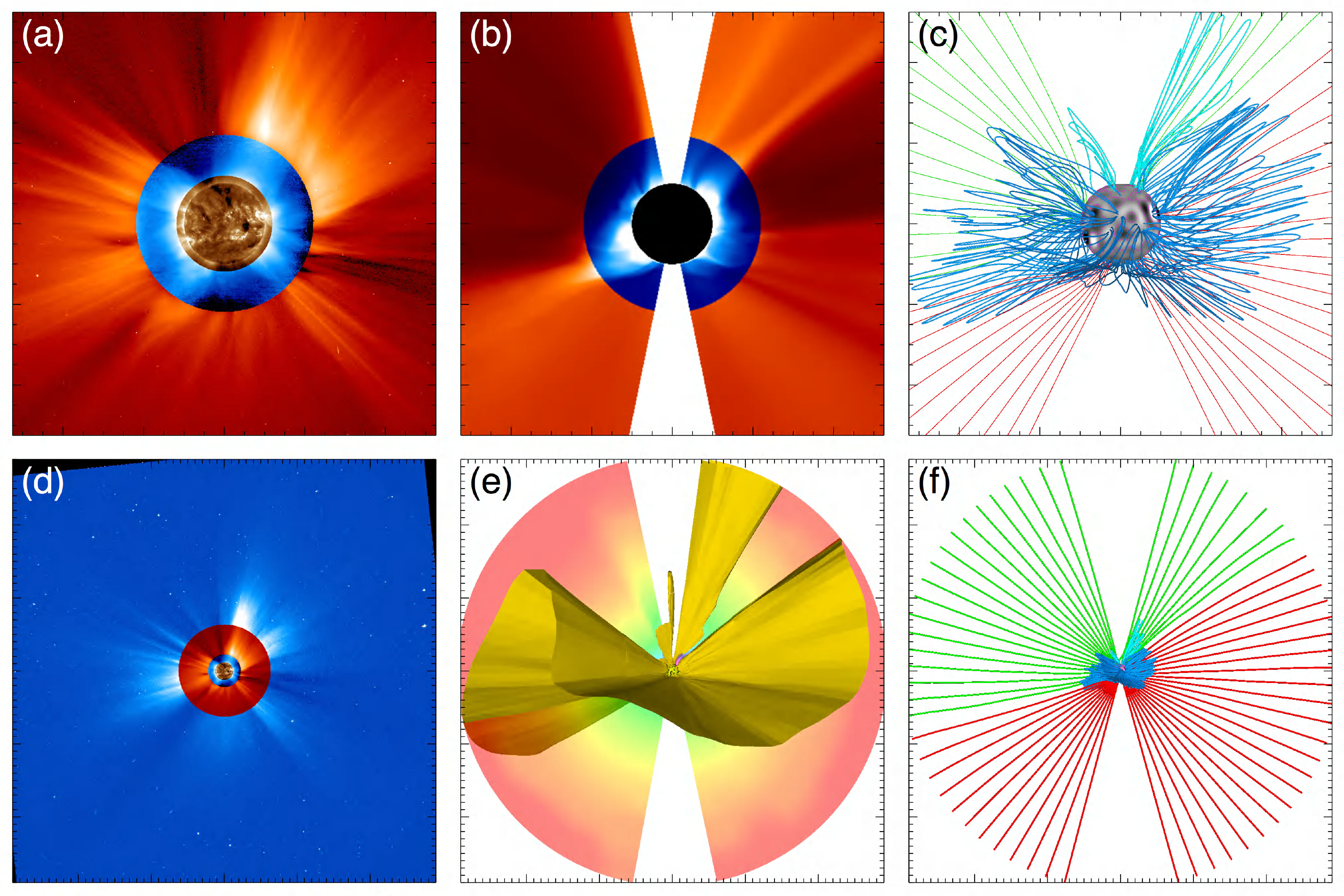} }
	\caption{(a): Composite image of the pre-eruption corona using EUV data from SDO/AIA in the 193~{\AA} channel and white-light data from the COSMO/K-Cor and SOHO/LASCO/C2 coronagraphs on July 9 around 19:00 UT. (b): Synthetic white-light intensity ratio image from ARMS simulation at $t=100$~hr. (c): Representative magnetic field lines showing the helmet streamer structure of the steady-state solar wind. Comparison to pre-eruptive white-light coronagraph structure. (d): Same as (a) with the addition of SOHO/LASCO/C3 data. (e): Visualization of the heliospheric current sheet as the $B_r=0$ isosurface. (f): Same as (c) for the LASCO/C3 field of view.}
	\label{f4}
\end{figure*}

We initialize the simulation magnetic field with the potential-field source-surface \citep[PFSS; e.g.,][]{Wang1992, Luhmann1998} reconstruction.
%
%
Figure~\ref{f3} shows magnetic field observations from the Global Oscillation Network Group \citep[GONG;][]{Harvey1996} of the National Solar Observatory (NSO) and two of the derived PFSS visualization data products associated with CR~2165 (top row) and the analogous plots depicting the initial $t=0$~hr magnetic field configuration of the ARMS simulation (bottom row). Panel~\ref{f3}(a) shows the NSO/GONG zero-point-corrected, daily updated $B_r$ synoptic map with the high-latitude filament-channel source region indicated by the yellow rectangle; July 9 is indicated by the vertical dashed line at $\phi=+60^{\circ}$. Panel~\ref{f3}(b) shows the NSO/GONG PFSS reconstruction of the global magnetic geometry including the positive (negative) coronal-hole open-field regions as green (red) areas, the extent of the helmet-streamer belt as the blue field line projections, and the $B_r=0$ contour at the $2.5\,R_\odot$ source surface indicating the location of the heliospheric current sheet (HCS) as the black line. Panel~\ref{f3}(c) shows the Earth view of the NSO/GONG PFSS configuration.
Panels \ref{f3}(d,e,f) show the same plots as above but for the simulation initial magnetic-field configuration, where we have used a truncated harmonic expansion ($\ell_{\rm max} = 25$) in the PFSS calculation. While the active-region-scale features are necessarily under-resolved in the ARMS version, the global-scale magnetic geometry is reproduced faithfully. The highest level of grid refinement is centered on the high-latitude filament-channel involved in the eruption.

We use the \citet{Parker1958} isothermal solar-wind model to construct the initial outflow conditions of our background solar wind. 
The number density, pressure, and temperature at the lower radial boundary are given by
$\rho_0/m_p = 2.59 \times 10^8$~cm$^{-3}$, 
$P_0 =0.10$~dyn~cm$^{-2}$, and 
$T_0 = 1.4\times10^6$~K, respectively.
Hence, the sound speed is $c_0 = 152$~km~s$^{-1}$ and the location of the critical point is $r_c = 4.10\,R_\odot$.  
The solar-wind speed at the outer boundary is $V_{\rm sw}(30R_\odot) = 410$~km~s$^{-1}$. 
We impose Parker's $V_{\rm sw}(r)$ profile at time $t=0$~hr and use it to set the initial mass-density profile $\rho(r)$ from the steady mass-flux condition ($\rho V_{\rm sw} r^2 =$~constant) throughout the computational domain. 
We then let the system relax until $t=100$~hr. The solar-wind relaxation process propagates the initial discontinuities in the PFSS solution at the source surface out of the domain and allows the amount of open and closed flux to adjust to the new pressure balance associated with the background outflow. 
The inner $r$ boundary conditions allow for mass flux into the domain where the density gradients develop, including into the open-field regions, thus providing the solar wind material. 

%

Our model wind and pre-eruption corona are compared with the white-light coronagraph observations in Figure~\ref{f4}. Figure~\ref{f4}(a) is a composite of the observational pre-eruption data, combining a LASCO/C2 image from 2015 July 9 at 19:00 UT, data from the K-Coronagraph (K-Cor) part of the Coronal Solar Magnetism Observatory (COSMO) taken at Mauna Loa on 2015 July 9 at 18:44~UT, and SDO/AIA 193~{\AA} data from 2015 July 9 at 19:00~UT. Panel~\ref{f4}(b) shows the synthetic white-light brightness ratio image constructed from the ARMS simulation data at $t=100$~hr. The total brightness ratio $I(t)/I_0$ is calculated from the line-of-sight integration of the Thomson scattering \citep{Billings1966,Vourlidas2006}; the intensity $I_0$ represents the white-light brightness image of the initial, spherically symmetric mass-density profile. Panel~\ref{f4}(c) shows representative magnetic field lines from the same perspective as (b). The open field lines from each polarity are shown in red and green, and the closed-field streamer-belt field lines are shown in blue. Panel~\ref{f4}(d) shows the same observational data as in \ref{f4}(a), but including data from the LASCO outer coronagraph C3, in a 30\,$R_\odot$ field of view. Panel~\ref{f4}(e) plots an isosurface of $B_r=0$, visualizing the 3D structure of the HCS, and panel \ref{f4}(f) plots the same field lines as panel \ref{f4}(c) in the expanded LASCO/C3 field of view. We note that the magnetic-field structure and the highly warped streamer belt in the simulation show quite reasonable qualitative agreement with the coronagraph streamer structures.

\subsection{Filament Channel Energization} \label{subsec:flows}

The eruption originated in the strongly sheared magnetic field of the high-latitude filament channel. This shear, along with the associated magnetic free energy needed to power the CME, is absent from the minimum-energy PFSS configuration. To energize the field and create the filament channel, we employ a statistically averaged version of the helicity condensation model \citep{Antiochos2013}. The Sun is known to generate magnetic helicity in the corona that is predominantly negative (left-handed) in the northern hemisphere  and positive (right-handed) in the southern \citep[e.g.][]{Pevtsov2014}, through subtle but persistent mechanisms that remain poorly understood and challenging to observe directly. Both vortical convection at the photosphere and the global-scale distribution of twisted flux emerging into the corona from below may contribute to the hemispheric pattern of injected helicity, which subsequently is transported via magnetic reconnection to PILs of the magnetic field. ARMS simulations have demonstrated that this model forms filament-channel-like coronal structures \citep{Knizhnik2015,Knizhnik2017a,Knizhnik2017b} that, in spherical geometry, can erupt to generate CMEs and associated eruptive flares \citep{Dahlin2019a}. 
{The statistically averaged implementation of the helicity condensation model was developed to support long-duration, full-Sun studies of filament-channel evolution \citep{Mackay2014,Mackay2018} as well as to facilitate investigations of individual eruptions (\citealt{Dahlin2019b}; \citealt*{Dahlin2021}, in preparation) such as the event studied here.} 
This model, in which horizontal magnetic flux is injected directly into the low corona, preferentially adjacent to PILs, is called Statistical Injection of Condensed Helicity (STITCH).

The STITCH sheared-flux generation is calculated from

\begin{equation}
\label{eq:stitch}
    \frac{\partial\boldsymbol{B}_S}{\partial t} = - \frac{1}{\lambda}   
    \boldsymbol{\nabla} \boldsymbol{\times} 
    \left( \sum_{i=1}^{8} \zeta^{(i)} B_r \right) \boldsymbol{\hat{r}} \;\;, 
\end{equation}

\noindent where the parameter $\lambda = 5 \times 10^{8}$~cm is the vertical scale of the helicity injection, 
$\boldsymbol{B}_S=B_\theta\boldsymbol{\hat{\theta}} + B_\phi\boldsymbol{\hat{\phi}}$, and 

\begin{equation}
\label{eq:zeta}
    \zeta^{(i)}(B_r,\theta,\phi,t) = K_0 \; f_B(B_r) \; f_\theta \left(\theta\right) \; f_\phi \left(\phi\right) \; f_t(t)
\end{equation}

\noindent defines a set of spatial and temporal envelope functions with an amplitude coefficient $K_0$. The envelope functions smoothly ramp the helicity condensation rate to zero outside of the high-latitude filament channel and define the temporal extent of the energization phases used. The mathematical forms of $f_B(B_r)$, $f_\theta(\theta)$, $f_\phi(\phi)$, and $f_t(t)$ are given in Appendix \ref{appendix:driving} along with their respective parameter sets. One advantage to this formulation is that the radial field distribution on the lower boundary remains unchanged throughout the energization: hence, the baseline magnetic energy is fixed during this phase, so that all of the energy added is magnetic free energy available to drive an eruption. The coefficient $K_0$ has dimensions of diffusivity (length$^2$ time$^{-1}$); its magnitude is determined by the characteristic spatial and temporal scales of the underlying helicity-injecting processes.
{We refer the reader to Appendix~\ref{appendix:driving} here for the technical implementation, and to the appendix of  \citet{Mackay2014} for the derivation of the helicity injection formalism.} 

\begin{figure}
    \centering
	\includegraphics[width=0.475\textwidth]{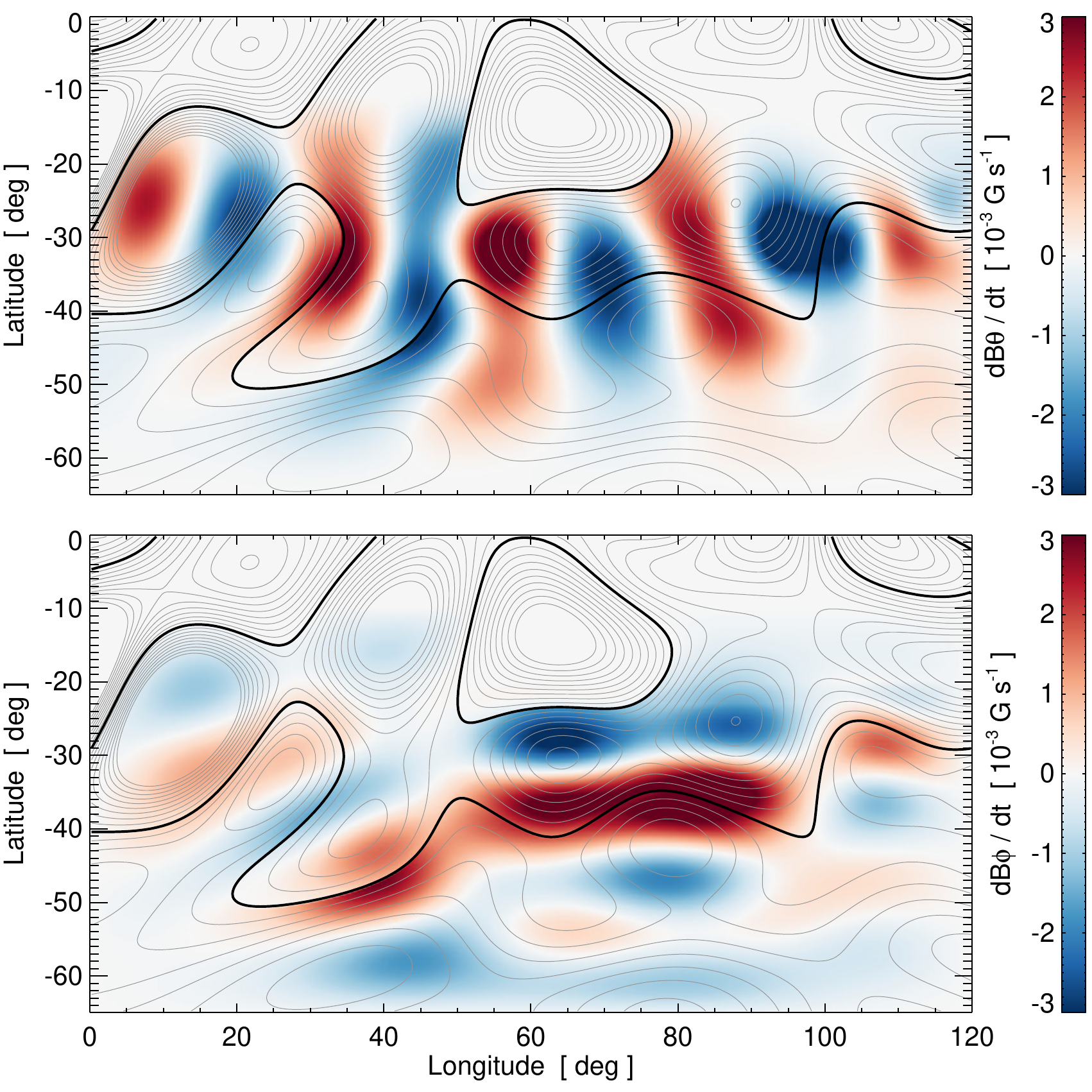}
	\caption{
	The imposed rates of change $\partial B_{\theta}/\partial t$ (top) and $\partial B_{\phi}/\partial t$ (bottom) during the STITCH horizontal-flux injection that energizes the filament channel.
	}
	\label{fstitch}
\end{figure}

The spatial distribution of the STITCH-injected shear is shown in Figure~\ref{fstitch} as the temporal derivatives of the horizontal field components given in Equation~(\ref{eq:stitch}). The top panel of Figure~\ref{fstitch} shows $\partial B_{\theta}/\partial t$, the bottom panel $\partial B_{\phi}/\partial t$, each at the maximum of the temporal envelope function ($f_t = 1.0$) during the energization phase. As can be seen in the figure, we stitched together eight separate patterns of flux injection in our simulation, so that shear was added more-or-less coherently all along the meandering PIL where the observed eruption occurred.

\begin{figure}[t!]
    \centering
	\includegraphics[width=0.475\textwidth]{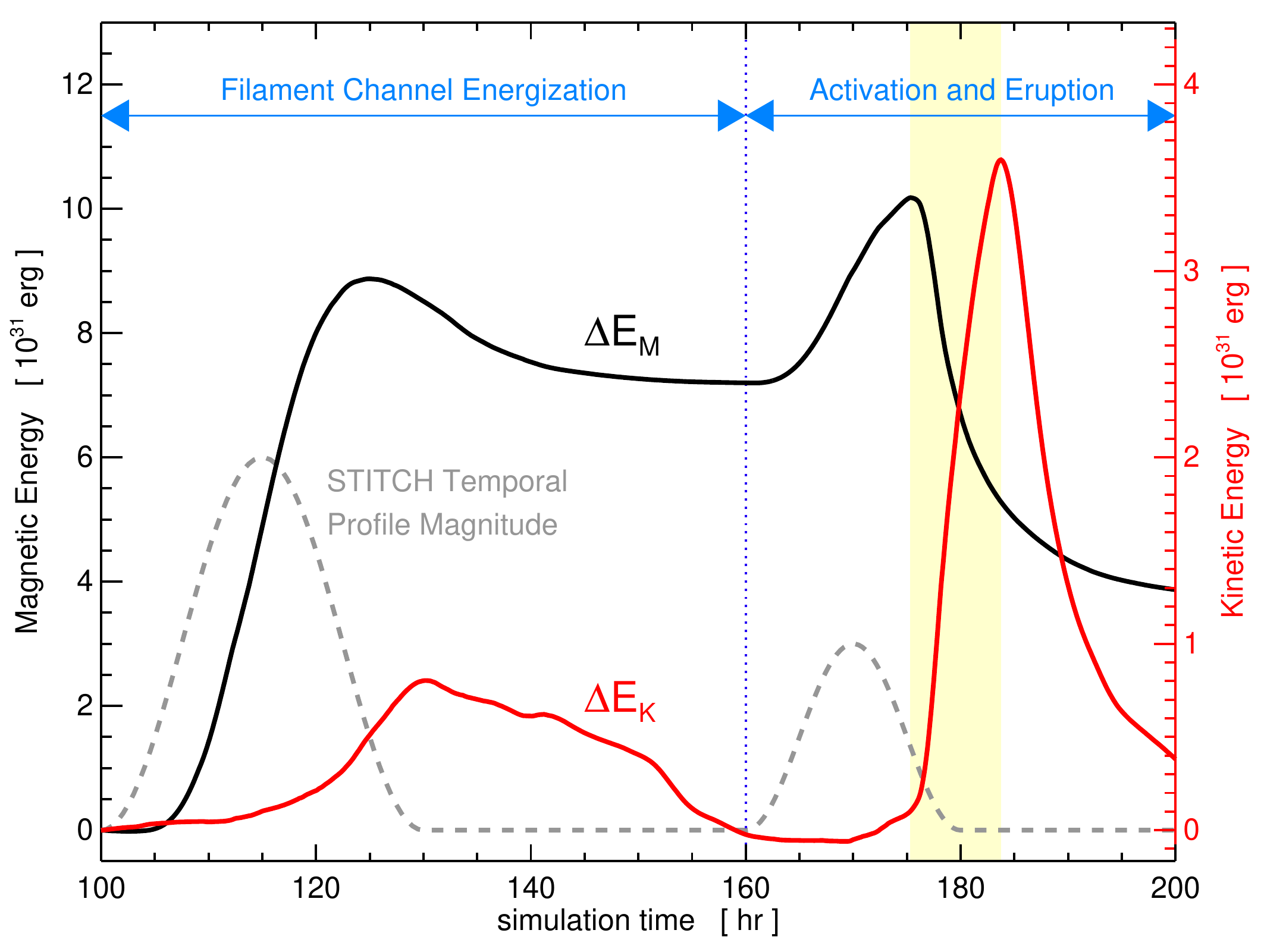}
	\caption{
	Evolution of the global magnetic ($\Delta E_M$) and kinetic ($\Delta E_K$) energies, together with the temporal profile of STITCH helicity injection, during the two main phases of the simulation: filament-channel energization and activation/eruption. {The yellow-shaded region corresponds to the impulsive phase of the eruptive flare.}
	}
	\label{feng}
\end{figure}

In this simulation, we employ two distinct energization phases in order to separate clearly the filament channel formation phase, which concentrates the sheared flux along the global PIL, from the activation and eruption phase, which transforms the system from a stable near-equilibrium state to an unstable, run-away eruption. 
Figure~\ref{feng} shows the global magnetic and kinetic energy evolution in our simulation through these two phases.
The energization phase lasts from 100~hr $\le t \le$ 160~hr, including 30~hr of STITCH generation of sheared flux with a smooth cosine temporal dependence for $t \in [100,130]$~hr followed by an additional 30~hr of relaxation to a new equilibrium of the large-scale sheared flux distribution above the filament channel. 
Figure~\ref{feng} plots the change in magnetic energy, $\Delta E_M(t) \equiv E_M(t) - E_M( 100{\rm \; hr} )$, in black and the change in kinetic energy, 
$\Delta E_K(t) \equiv E_K(t) - E_K( 100{\rm \; hr} )$, in red. The duration of the STITCH driving patterns are indicated with the gray dashed line; the vertical blue dotted line at $t=160$~hr separates the energization and eruption phases. The total magnetic energy accumulated at the end of the energization phase is $\Delta E_M(160{\rm \; hr}) = 7.20 \times 10^{31}$~erg.

%

\begin{figure*}
    \centering{
	\includegraphics[width=0.95\textwidth]{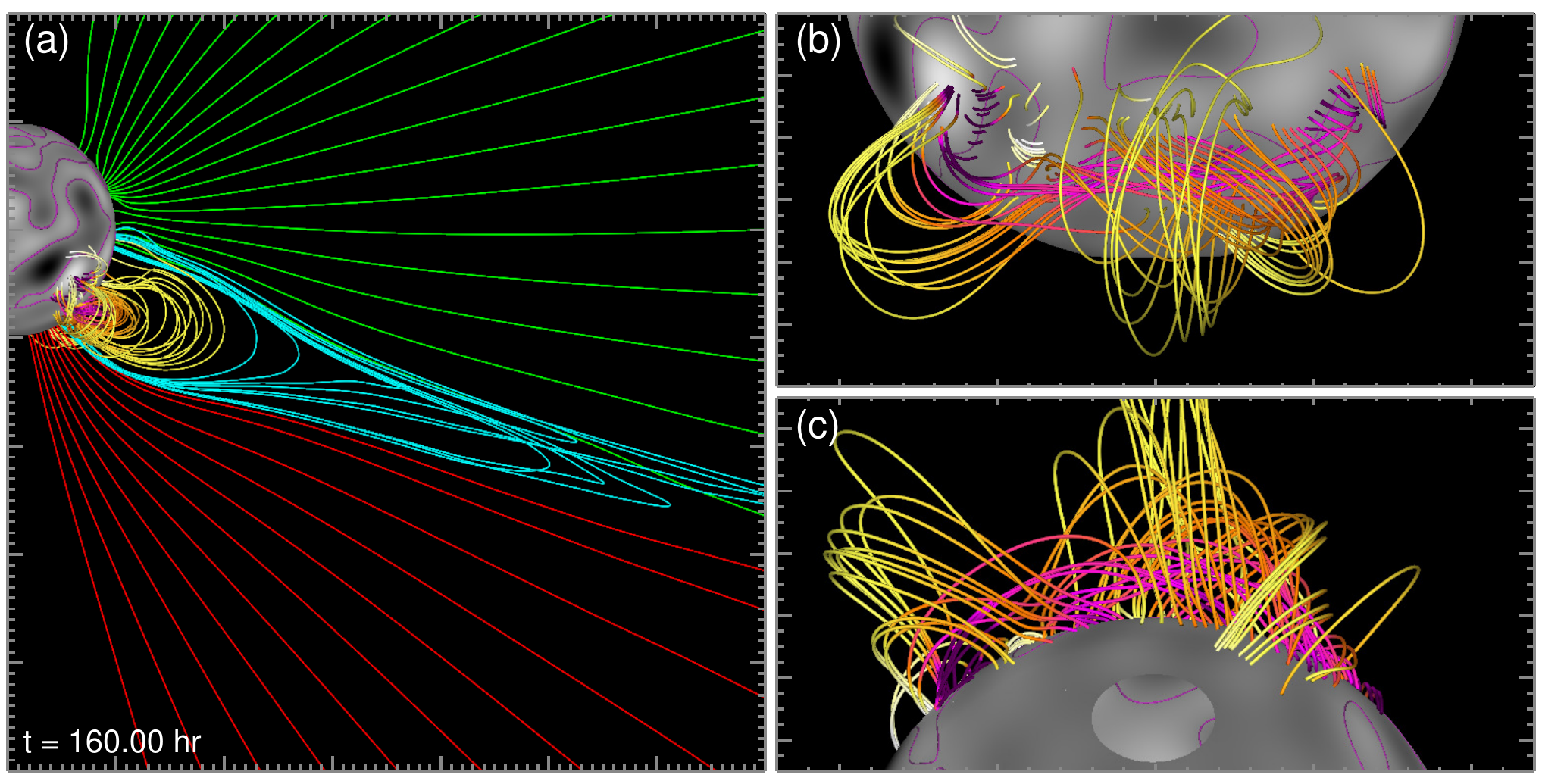}
	}
	\caption{Representative magnetic field lines illustrating the 3D structure of the filament channel. Field lines are colored by flux system at $t=160$~hr: positive (negative) polarity open fields are green (red), overlying streamer flux is cyan, and filament-channel field lines are colored magenta to yellow, proportional to $B_\phi$. (a): Viewpoint from a central meridian of $\phi=-30^{\circ}$ longitude so that the filament-channel center appears on the limb. (b): Filament-channel field lines from the `Earth view' of $\phi=60^{\circ}$. (c): View from above the south pole so that the filament channel appears on the limb, approximately parallel to the plane of the sky. The animated version of this figure runs from $t=160$~hr to $t=190$~hr and shows the whole filament-channel activation and eruption process.\\
	(An animation of this figure is available.) }
	\label{fflside}
\end{figure*}

Figure~\ref{fflside} shows representative magnetic field lines that illustrate the structure of the sheared filament channel and the overlying flux systems at the end of the energization phase. Figure~\ref{fflside}(a) shows the high-latitude filament channel (magenta-to-yellow field lines) on the west limb along with the overlying streamer flux (light blue) and the positive (negative) open fields in green (red). Figures~\ref{fflside}(b) and \ref{fflside}(c) show two close-up perspectives of the sheared filament-channel field lines shown in panel (a). The magenta-to-yellow color scale represents the $B_\phi$ component: the most sheared field lines with the largest $B_\phi$ values are magenta, the least sheared are yellow.  
We note that the field lines develop a weak twist from the structure of the helicity injection acting on the global PIL. The long horizontal field lines above the PIL form the characteristic dips found in many prominence-field models and observations \citep[e.g.,][]{DeVore2000b,Parenti2014,Gibson2018,Patsourakos2020}. 

\section{Modeling the CME Eruption} \label{sec:simresults}

\subsection{CME Initiation and Filament Eruption Dynamics} \label{subsec:filament}

The evolution of the total magnetic and kinetic energies during the activation and eruption phase, $t>160$ hr, are shown in Figure~\ref{feng}. Activation is achieved through another period of STITCH energization, this time for a shorter duration (20~hr) and a lower magnitude (50\% of the first period) compared to the earlier filament energization phase. The STITCH profiles are depicted by the height of the dashed gray line in Figure~\ref{feng}. During the activation phase, enough additional magnetic energy accumulates to facilitate the transition from stable, slow evolution to an unstable, runaway eruption.  

{The eruption occurs due to the evolution of the stressed magnetic field and the interaction between the energized filament channel flux and the large-scale, overlying and adjacent coronal field configurations \citep[e.g. see description in][]{Lynch2016b}. While the STITCH energization does not directly impose a velocity field at the lower boundary, the imposed electric field can be thought of as generating an ``effective'' foot point displacement with a velocity $v_{\rm eff}$. Examining the magnitudes of the velocity field at $r=1.05 R_\odot$ along the filament channel region, we find $v_{\rm eff} \le 20  {\rm \; km \; s}^{-1}$. Since the Alfv\'{e}n speed, $v_A$, in the same region ranged from $500-1000$~km~s$^{-1}$, $v_{\rm eff}$ at the peak of the helicity injection rate is $\sim$13\% of the sound speed and on the order of $\sim$4\% of the Alfv\'{e}n speed, i.e. $v_{\rm eff}/v_A < v_{\rm eff}/c_0 << 1$. Therefore, our STITCH energization can be considered quasi-static evolution rather than providing any sort of ``direct driving'' of the eruption.}

As in the analysis of our previous CME simulations \citep[e.g.][]{Lynch2013,Lynch2019}, we define the so-called ``impulsive phase'' of the eruption as the interval in which a sharp rise in kinetic energy, $\Delta E_K$, coincides with a similarly sharp drop in magnetic energy, $\Delta E_M$. 
The rapid conversion of magnetic to kinetic energy has been shown to be an excellent indication of the flare current sheet transitioning to fast magnetic reconnection, which is accompanied by the formation and ejection of plasmoids to process the significantly increased transfer of magnetic flux and mass \citep{Karpen2012,Lynch2016a,Lynch2019,Dahlin2019a}.    

In our simulation, the maximum $\Delta E_M(t_M^*) = 1.02 \times 10^{32}$~erg occurs at $t_M^* = 175.33$~hr, and the maximum $\Delta E_K(t_K^*) = 3.60 \times 10^{31}$~erg at $t_K^* = 183.75$~hr.
Thus, the impulsive phase corresponds to the time interval $t_M^* \le t \le t_K^*$, shaded yellow in Figure~\ref{feng}.
%
%
%
%
The overall ratio of magnetic-to-kinetic energy conversion during this impulsive phase is 

\begin{equation}
\frac{\Delta E_K(t_K^*)-\Delta E_K(t_M^*)}{\Delta E_M(t_M^*)-\Delta E_M(t_K^*)} = \frac{3.494 \times 10^{31}}{4.896 \times 10^{31}} = 71.4\% \; .
\end{equation}

\noindent {Only a small portion of the remaining free magnetic energy released during the impulsive phase is captured in our simulation. There is a modest increase in the total internal energy of $2.58 \times 10^{30}$~erg (5.3\% of the magnetic energy release) associated with adiabatic compression. The remaining 23.3\% of the free energy released is lost in our isothermal model but would correspond to bulk plasma heating through magnetic and viscous dissipation, enhanced radiation output, and energetic particle acceleration.} 
We note that the total drop in magnetic energy between its maximum at $t_M^*$ and the end of our simulation ($t_f=200$~hr) is $\Delta E_M(t_M^*) - \Delta E_M(t_f) = 6.31 \times 10^{31}$~erg.

%
%
%

%

The animation of Figure~\ref{fflside} illustrates the magnetic-field evolution in the low corona during the filament activation ($160$~hr $\le t \lesssim 175$~hr) and its eruption resulting in the CME ($t \gtrsim 175$~hr). During the activation phase, the system evolution has two primary features. First, the segment of the helmet-streamer belt above the extended filament channel inflates both laterally and radially. This gradual swelling before an eruption is a common feature of slow streamer blowout CMEs \citep[e.g.,][]{Sheeley1997, Vourlidas2018}. It can be most easily seen in Figure~\ref{fflside}(a) as the overlying streamer cyan field lines and then some of the outermost yellow field lines near the east limb stretching out and ``opening up'' beyond the edge of the plot window at $r \sim 7R_\odot$ \citep[see also][]{Lynch2016b}. Second, the filament-channel sheared fields rise gradually and asymmetrically, with the eastern limb (left-hand side of panels b,c) expanding first and the expansion sweeping from east to west along the PIL. The expanding magenta field lines become orange and then red with height indicating the radial fall-off of the $B_\phi$ magnitude. By $t=176.67$~hr, about half of the filament-channel field lines have opened up and the eruptive-flare reconnection is well underway. The eruption proceeds rapidly thereafter, again sweeping from east to west. By $t=179.0$~hr, all of the yellow outer filament-channel field lines have opened from the perspective of panels (b,c), whereas in panel (a) one sees the CME in the midst of the eruption having rapidly expanded and acquired a complex, twisted structure that mixes red, green, cyan, and yellow field lines. For $t > 179.0$~hr, the yellow field lines reconnect in the flare current sheet and start to close back down as post-eruption arcade loops. All have reconnected by $t = 184.67$~hr. Plasmoids form in the eruptive-flare current sheet to facilitate the rapid flux transfer and can be seen trailing the eruption \citep[e.g., see frame at $t=188.67$~hr; also][]{Riley2007,Webb2016,Chae2017}.

{It is worth pointing out that the post-eruption filament channel fields do not return to the exact configuration of the pre-energization state. There is still $\Delta E_M(t_f) = 3.87 \times 10^{31}$~erg worth of free magnetic energy stored in the shear and twist in the magnetic fields above the high-latitude PIL after the eruption. This is seen in the $\Delta E_M$ energy curve of Figure~\ref{feng} and is visible at the end of the Figure~\ref{fflside} animation in the lowest-lying field lines. This is essentially a universal feature in both numerical simulations and the observations. The erupting flux does not ``open up'' all the way to the PIL, rather there remains a comparatively small but non-trivial component of the sheared field structure of the filament channel.} 

\subsection{Flare Ribbons, EUV Dimmings, and the Post-Eruption Arcade} \label{subsec:arcade}

\begin{figure*}[t!]
	\centering{
	\includegraphics[width=0.95\textwidth]{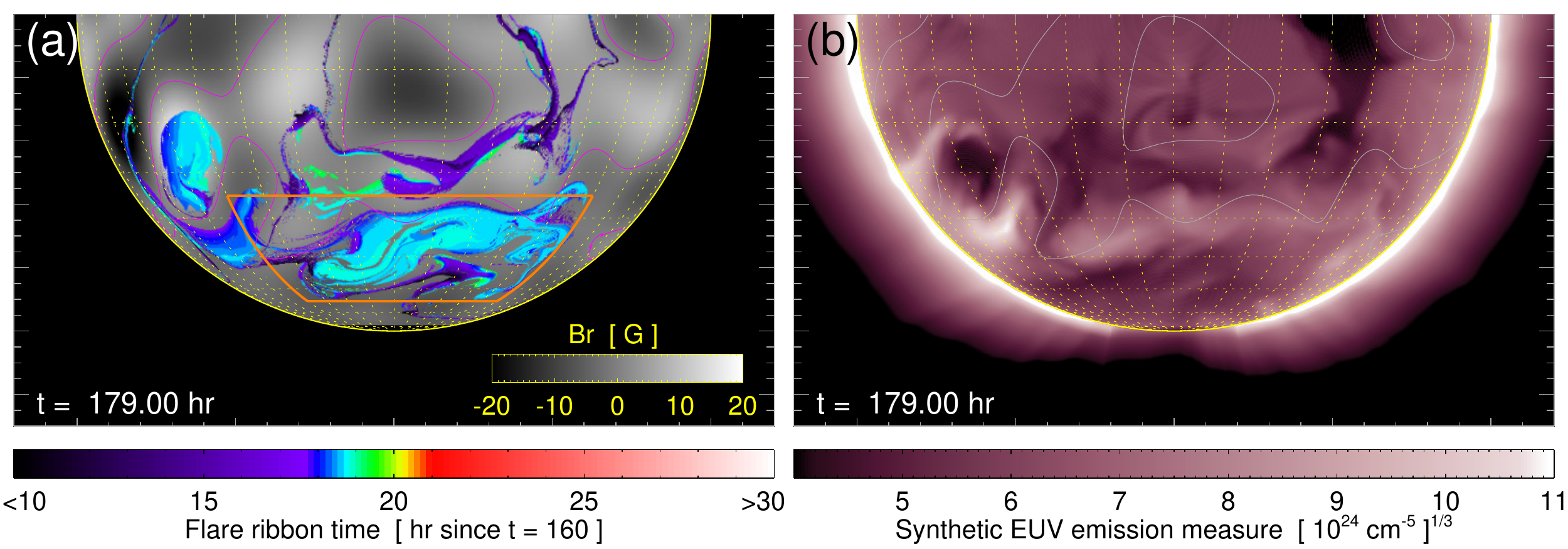}
	}
	\caption{(a): Snapshot at $t=178.33$~hr of the spatiotemporal evolution of the flare ribbons sweeping out the reconnection flux plotted over the $B_r$ distribution. Magenta lines show the PILs. The orange lines depict the sub-region of the reconnection flux used in the calculation of $\Phi_{\rm rxn}^{\rm ROI}$ (see text for details). (b): Synthetic EUV {emission measure} showing the evolution of the coronal dimming regions and the development of the post-eruption arcade. The animated version of this figure runs from $t=160.33$~hr to $t=200$~hr and shows the full development of the flare ribbons, dimmings, and post-eruption arcade.\\
	(An animation of this figure is available.) }
	\label{feuv}
\end{figure*}

To track the evolution of the reconnection ribbons in the simulation data, we use the \citet{Lynch2019} implementation of the \citet{Kazachenko2017} procedure for analyzing two-ribbon flares in SDO data. The change in field-line length $L$ between successive frames is calculated as $\Delta L = L(t) - L(t-\Delta t)$; we use this as a proxy for the rapid geometric reconfiguration of the magnetic field line connectivity. Our simulation output frames are in $\Delta t = 20$~min intervals, and we track a $768 \times 1568$ uniform grid of field lines at the lower radial boundary in $(\theta, \phi)$ over the region $\theta \in{[-78.5^\circ,11.5^\circ]}$ and $\phi \in{[-30^\circ,150^\circ]}$.
If a field line becomes shorter by $\Delta L \le -2 R_\odot$ over $\Delta t$ and both footpoints are connected to the lower $r=R_\odot$ boundary, then we consider that field line (pixel) to have undergone reconnection as either an open field line closing down or a closed field line becoming significantly shorter. These reconnection pixels are accumulated in time to create the cumulative ribbon-area map.

Figure~\ref{feuv}(a) shows the evolution of the area swept out by reconnection at $t=178.33$~hr, during the impulsive phase of the eruption, over the $B_r$ distribution at the $r=R_\odot$ boundary from the viewpoint of the SDO observations in Figure~\ref{f1}. The color scale indicates the time when the magnetic flux bundle at a given pixel first reconnects. The large-scale evolution of the ribbon morphology shows qualitative agreement with the generic, universal picture of two-ribbon-flare evolution, in which the ribbons first grow rapidly parallel to the eruption-associated PIL and then expand away from the PIL in the perpendicular direction more slowly \citep[the so-called zipper effect;][]{Moore2001, Linton2009, Qiu2009, Aulanier2012, Priest2017}. 

The MHD simulation's large-scale flare-ribbon evolution also shows qualitative agreement with the SDO/AIA observations of the low corona for the 2015 July 9--10 filament eruption. As described in $\S$\ref{subsec:disk_obs}, multiple wavelengths showed an overall east-limb to west-limb evolution of the various eruption signatures on the disk. In Figure~\ref{feuv}(a), we see precisely that temporal evolution from east to west in the cumulative ribbon-area map in the impulsive phase. 

For a more direct comparison to the SDO/AIA observations, we calculate the time-dependent synthetic EUV {emission measure} from the simulation's evolving density distribution.
The synthetic EUV {emission measure} is modeled as $I_N(x,y) = \int dz' \; n_e^2(x,y,z')$, where we take $n_e = n_p = \rho/m_p$. We constructed an image from a $512 \times 512$ 2D array of lines-of-sight covering the range
$-1.2R_\odot \le x \le 1.2R_\odot$,
$-1.3R_\odot \le y \le 0.12R_\odot$ and have used 768 samples along the $z'$ direction ($-2R_\odot \le z' \le 2R_\odot$) for the integration of the $n_e^2$ values.
Figure~\ref{feuv}(b) shows the resulting synthetic EUV {emission measure} at the same time ($t=178.33$~hr) as the flare ribbon area map in \ref{feuv}(a). 
We plot the cube-root of the calculated synthetic EUV intensity, i.e. $(I_N)^{1/3}$, to compensate for the large dynamic range and use the AIA 211~{\AA} color map to facilitate comparison with the corresponding features of Figure~\ref{f1}(b).
Here we see both the dynamic formation and evolution of dimmings associated with the flux-rope foot points and the rapid evacuation of coronal material above the filament channel during the eruption. The latter EUV dimmings are followed by an emission enhancement from the post-eruption arcade (PEA) as the evacuated filament channel refills with coronal material. Here again, we note the east-to-west development of the synthetic EUV features. An animation of this figure is provided in the online version of the article. 

\begin{figure}[!t]
    \centering
	\includegraphics[width=0.475\textwidth]{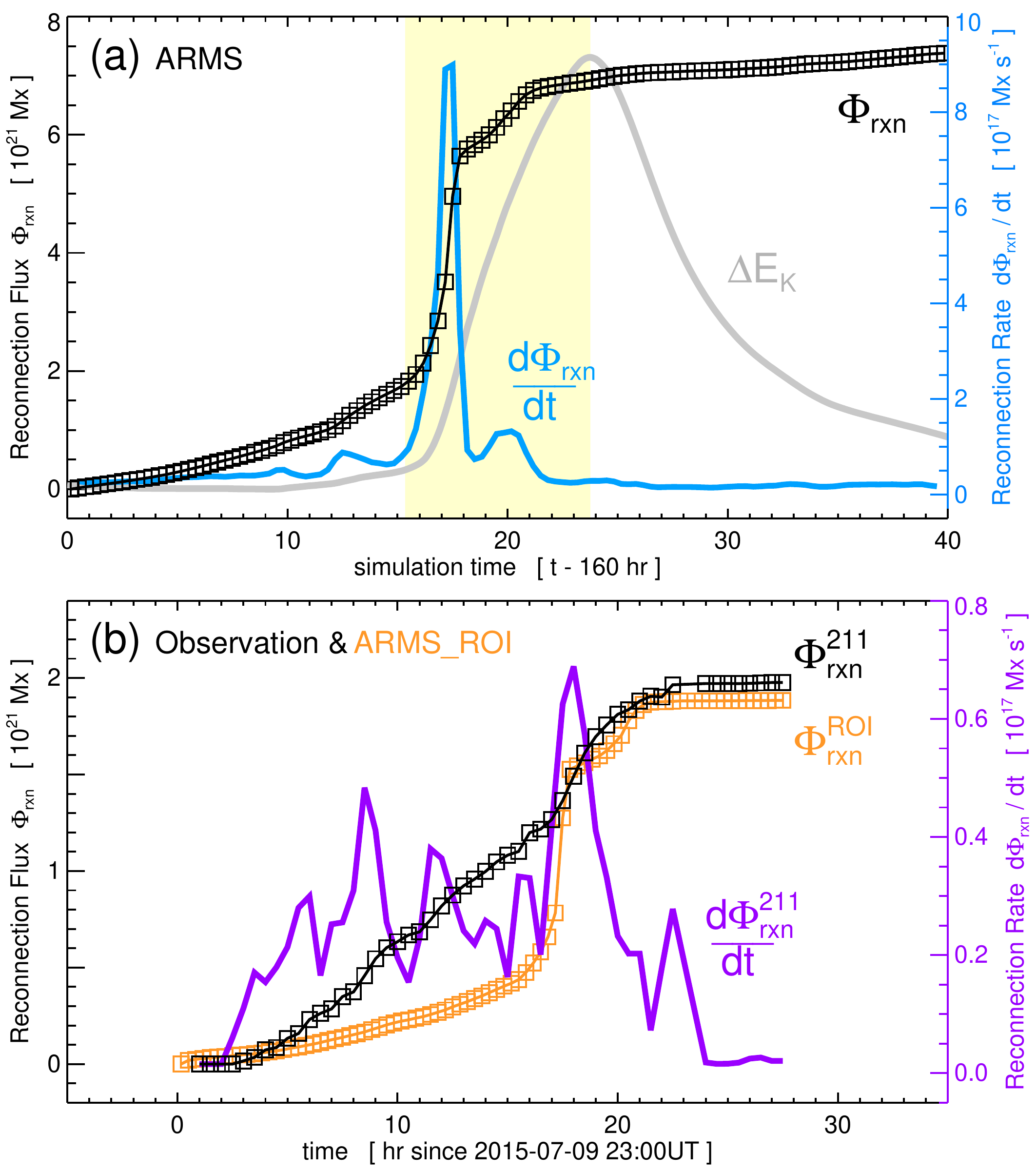}
	\caption{(a): ARMS simulation reconnection flux $\Phi_{\rm rxn}$ and reconnection rate $d\Phi_{\rm rxn}/dt$. {The yellow-shaded region corresponds to the impulsive phase of the eruptive flare.} (b): Reconnection flux and rate ($\Phi_{\rm rxn}^{211}$, $d\Phi_{\rm rxn}^{211}/dt$) estimated from the SDO HMI and running-difference AIA 211\AA\ observations over the course of the eruption. The $\Phi_{\rm rxn}^{\rm ROI}$ reconnection flux is from the ARMS simulation in a $30^\circ \times 90^\circ$ in $\theta, \phi$ region of interest (ROI) centered on the negative flux side of the filament channel.}
	\label{feuv2}
\end{figure}

The normal magnetic flux swept by the ribbon area on the photosphere corresponds to the flux processed by the flare reconnection in the corona \citep{Forbes1983}. We calculate the simulation reconnection flux as

\begin{equation}
\Phi_{\rm rxn}(t) = \onehalf \int dA \;\;  \left[ |B_r^{(+)}| + |B_r^{(-)}| \right],
\end{equation}

\noindent where $A(t)$ is the cumulative ribbon area shown in Figure~\ref{feuv}(a) and $|B_r^{(+)}|$, $|B_r^{(-)}|$ are the magnitudes of the positive and negative polarities of the radial-field distribution at $r=R_\odot$.
Figure~\ref{feuv2}(a) plots the temporal evolution of $\Phi_{\rm rxn}$ (black) and the reconnection rate $d\Phi_{\rm rxn}/dt$ (blue). The kinetic-energy curve from Figure~\ref{feng} is shown in gray to illustrate that the onset of fast reconnection occurs prior to the global kinetic-energy rise.

Figure~\ref{feuv2}(b) plots an estimate of the observed reconnection-flux profile, $\Phi_{\rm rxn}^{211}$, from the SDO/AIA and HMI data for the 2015 July 9--10 filament eruption. The observational profile uses the sequence of HMI $B_{\rm LOS}$ magnetograms and the PEA area estimate based on the signatures in running-difference 211~{\AA}, i.e.\ the dimming and refilling of the emission above the PIL. We note that, due to the complexity of AR emission structures north of the high-latitude PIL, our $\Phi_{\rm rxn}^{211}$ estimate includes only the negative-polarity flux south of the PIL. An equal amount of positive-polarity flux must participate, but it is not unambiguously distinguishable from the AR flux to the north, so we omit it from our calculation. Appendix~\ref{appendix:obsflare} describes the procedure for the observational thresholding used to create the cumulative flare-ribbon maps from the SDO/AIA data. 

Figure~\ref{feuv2}(b) also plots a portion of the total ARMS reconnection flux, $\Phi_{\rm rxn}^{\rm ROI}$, calculated from a region of interest (ROI) defined by $\theta \in [125^{\circ}, 155^{\circ}]$ (corresponding to a latitude range of $[-35^{\circ}, -65^{\circ}]$) and $\phi \in [20^\circ, 110^\circ]$. Here we use the cumulative ribbon area shown in Figure~\ref{feuv}(a) but only with the magnitude of the negative-polarity flux south of the PIL, just as in the observational estimate.

There are both similarities and differences between the model reconnection flux and the observational estimate. For example, the total reconnection flux at the end of the MHD simulation, $t_f = 200$~hr, is $\Phi_{\rm rxn}(t_f) = 7.38 \times 10^{21}$~Mx, whereas the analogous quantity estimated from the AIA observations is $\Phi_{\rm rxn}^{211}(t_{{\rm obs},f}) = 1.98 \times 10^{21}$~Mx, where we have taken $t_{{\rm obs},f}$ as 2015 July 11 at 03:00 UT.
The value of the ROI reconnection flux is almost identical to the observational estimate, $\Phi_{\rm rxn}^{\rm ROI}(t_f) = 1.96 \times 10^{21}$~Mx.
The maximum ARMS simulation reconnection rate obtained for both the full pixel map and the ROI portion is $d\Phi_{\rm rxn}/dt = 8.87 \times 10^{17}$~Mx~s$^{-1}$, whereas observationally we estimate $d\Phi_{\rm rxn}^{211}/dt = 6.86 \times 10^{16}$~Mx~s$^{-1}$.
The ratio between the simulated and the observational reconnection fluxes is $\Phi_{\rm rxn}/\Phi_{\rm rxn}^{211} \approx 3.7$ with the total reconnection area, but within the ROI area it is $\Phi_{\rm rxn}^{\rm ROI}/\Phi_{\rm rxn}^{211} \approx 0.99$.
The reconnection-rate ratio is $(d\Phi_{\rm rxn}/dt)/(d\Phi_{\rm rxn}^{211}/dt) \approx 12.9$ for both the full ribbon area and the ROI area. 

The Figure~\ref{feuv2}(b) comparison between $\Phi_{\rm rxn}^{\rm ROI}$ and $\Phi_{\rm rxn}^{211}$ makes clear that both reconnection-flux profiles have a gradual, linear increase preceding the largest increase in reconnection rate, which occurs at $t_{\rm sim} \approx 176$~hr and $t_{\rm obs} \approx$ 2015 July 10 at 16:00 UT, respectively. The slope during the linear increase is greater in the observations, whereas the ARMS eruption has a more typically characteristic ``impulsive'' phase than is readily identifiable in the observations. We note that the duration before reaching the maximum reconnection flux is approximately 17~hrs in both cases. 
Despite the quantitative differences in the reconnection profiles, there is substantial qualitative agreement between the simulation results and the observational estimate of the regions on the disk face associated with the reconnection flux. 

Using the \citet{Kazachenko2017} power-law scaling relation between total unsigned reconnection flux and flare strength/classification from X-ray emission, our simulation $\Phi_{\rm rxn}$ corresponds to an unsigned flux of $1.48 \times 10^{22}$~Mx and an X1.1 class flare, while the observational estimates from 211~{\AA} correspond to $3.95 \times 10^{21}$~Mx and an M1.6 class flare. 
The observed 1--8~{\AA} GOES X-ray flux, however, never exceeded C1.9 between 2015 July~9--11\footnote{See \url{ftp://ftp.swpc.noaa.gov/pub/warehouse/2015/WeeklyPDF/prf2080.pdf}.}. The NOAA flare most relevant to our high-latitude filament eruption was a B7.3 flare on 2015 July 10, peaking at 01:26UT, from AR12384 when it was positioned at S22E54. All three of the C-class flares during this period were from regions in the Northern hemisphere.
The lack of observed X-ray emission is consistent with the \citet{Lynch2016b} interpretation for a similarly large CME source region, whose spatial and temporal scales were of order $R_\odot$ and 24~hr, respectively. The Poynting flux associated with the reconnection flux swept into the flare current sheet provides an upper limit on the free magnetic energy available for conversion into bulk plasma heating over the flare arcade volume. This energy is spread over such a large area at such a slow rate in this event that the resulting energy flux into the newly formed flare-arcade loops is insufficient to generate significant plasma heating and attendant X-ray emission.

It is also instructive to compare our results to the magnetic-flux estimates for other large quiet-Sun filament and/or polar-crown prominence eruptions because properties of the magnetic field (handedness, orientation, and magnetic flux content) are important quantities for making the CME--ICME connection \citep[e.g.][]{Demoulin2008, Palmerio2017, Gopalswamy2018}.
In general, comparisons between reconnection flux and magnetic-cloud (MC) poloidal/twist flux $\Phi_p^{\rm MC}$ measured in situ have shown robust correlations with $\Phi_{\rm rxn} \gtrsim \Phi_p^{\rm MC}$ \citep[e.g.,][]{Qiu2007, Kazachenko2012, Hu2014, Gopalswamy2017}.
Recently, \citet{Cliver2019} calculated the reconnection flux during a large, quiet-Sun filament eruption on disk, obtaining $3.2 \times 10^{21}$~Mx, very similar to our observational estimate here.
Reconnection fluxes of this order of magnitude, a few times $10^{21}$ Mx, are comfortably within the distribution of both SDO/AIA flare reconnection fluxes \citep[$10^{20\rm{-}22}$ Mx;][]{Kazachenko2017} and in-situ MC ICME poloidal/twist flux estimates \citep[$10^{21\rm{-}22}$~Mx;][]{Lynch2005}.

\begin{figure*}[!t]
    \centering{
    \includegraphics[width=0.95\textwidth]{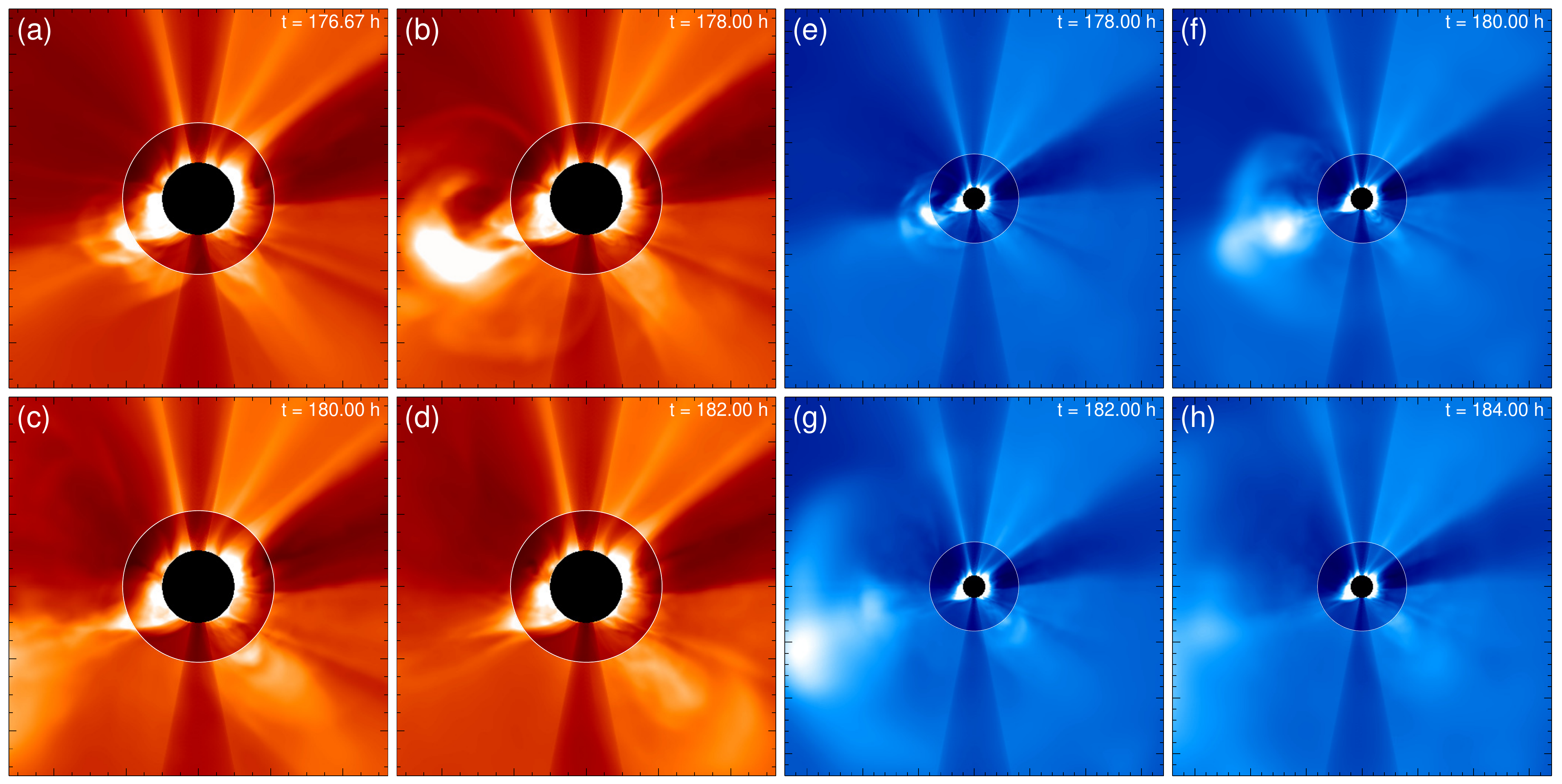}
    }
	\caption{Synthetic white-light coronagraph images during the multi-stage eruption in the style of Figure~\ref{f2}. (a)--(d): Simulated LASCO/C2 field of view. (e)--(h): Simulated LASCO/C3 field of view. The animated version of this figure runs from $t=160$~hr to $t=191$~hr and shows the whole passage of the CME through the fields of view of both synthetic coronagraphs.\\
	(An animation of this figure is available.)  }
	\label{fwlsim}
\end{figure*}

\subsection{White-Light Morphology, CME Structure, and Kinematics} \label{subsec:coronagraph}

\begin{figure*}
	\centering{
	\includegraphics[width=0.95\textwidth]{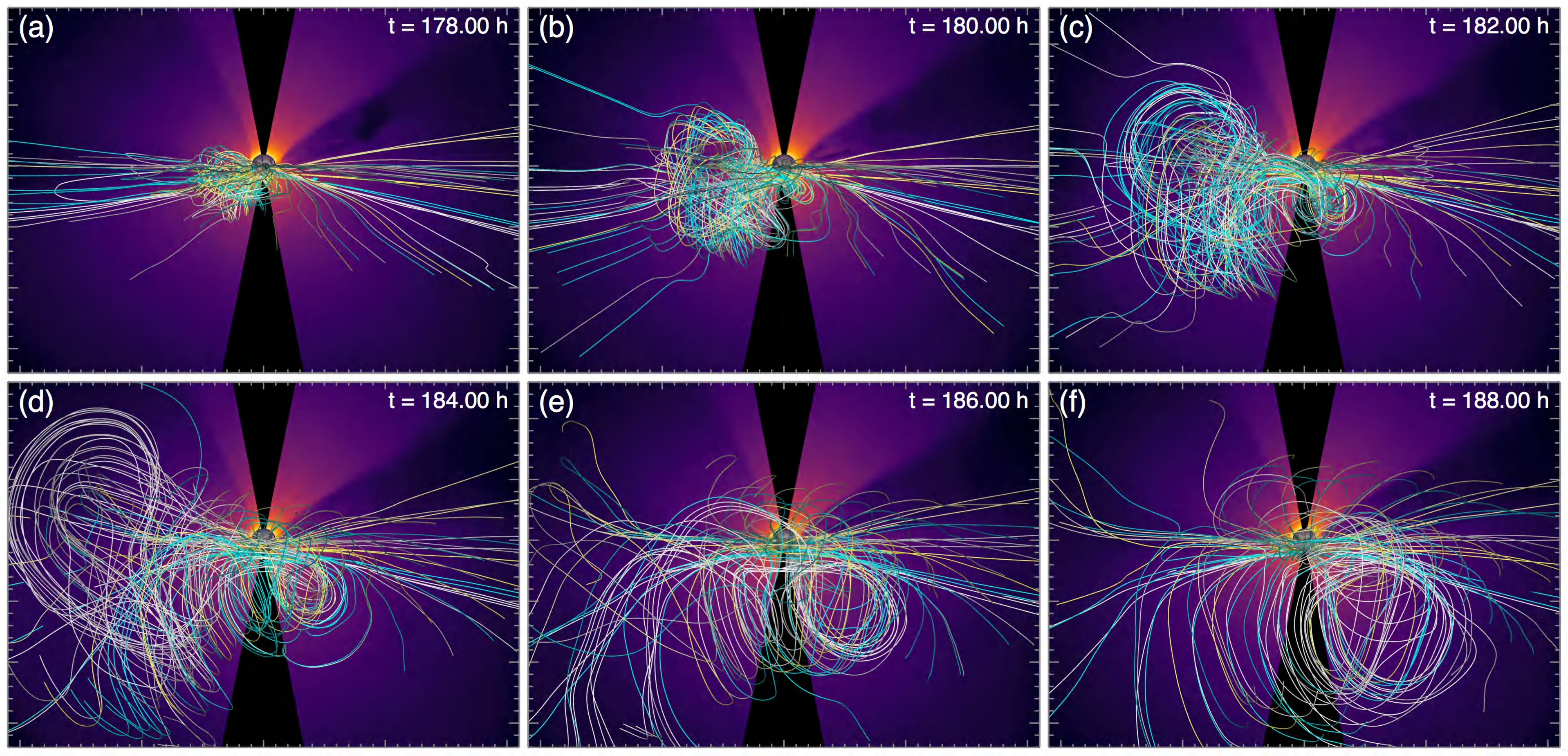}
	}
	\caption{Representative field lines illustrating the CME's 3D magnetic structure. The log-scale mass density is shown in the plane of the sky and the image perspective is from Earth's view. Panels (a)--(d) are the same simulation times shown of the synthetic LASCO/C3 panels in Figure~\ref{fwlsim}(e)--(h). The animated version of this figure runs from $t=160$~hr to $t=200$~hr and shows the full development of the magnetic field lines during the eruption process.\\
	(An animation of this figure is available.) }
	\label{fprop}
\end{figure*}

{To compare the simulation results with the coronagraph observations of the CME's evolution ($\S$\ref{subsec:corona_obs}), we constructed synthetic white-light images for the entire filament-activation and CME-eruption phase.} The procedure was outlined in $\S$\ref{subsec:init}.
Figures \ref{fwlsim}(a-d) show the resulting C2-like field of view ($\pm 5.25 R_\odot$ in $x,y$) for comparison with Figures \ref{f2}(a-d); \ref{fwlsim}(e-h) show the C3-like field of view ($\pm 17 R_\odot$ in $x,y$) for comparison with \ref{f2}(e-h). In each of the panels, we indicate the position of the C2 and C3 occulting disks at $2.1\,R_\odot$ and $4\,R_\odot$, respectively. 
An animated version of this figure is available in the online article.

The main features of the simulation's white-light imagery are the following. The east- and west-limb portions of the extended CME eruption are separated enough in space and time that they could easily be interpreted as two separate events. This is exactly how they are described in the CDAW LASCO CME Catalog\footnote{Available at \url{https://cdaw.gsfc.nasa.gov/CME_list/}. See entries 2015/07/09~19:00:05 (position angle 145$^\circ$, width 145$^\circ$) and 2015/07/10~02:24:04 (position angle 181$^\circ$, width 138$^\circ$).}.
%
%
%
%
%
The east-limb component of the CME features the classic ``three-part'' structure \citep[e.g.,][]{Illing1985,Vourlidas2013} consisting of an enhanced leading edge, dark cavity, and bright core (cf. Figure~\ref{f2}(a,e)). The west-limb component looks more like a classic ``loop'' CME without a discernible core (cf. Figure~\ref{f2}(d,h)). Because the western portion erupts in front of a background streamer (in both observations and simulation), one might surmise that there could be an ejecta-related `core' enhancement. However, the animations of Figures~\ref{f2} and \ref{fwlsim} make clear that this is not the case.

The comparison between the white-light coronagraph imagery and the actual magnetic-field structure of the CME ejecta can be done with the simulation results in a way that is simply not possible with the observational data. Fortunately, insights obtained from examining the relationship between features of the simulation's magnetic field and synthetic white-light morphology can be applied to the interpretation of the observational data, as well. 
  
The animation of Figure~\ref{fflside}(a) shows that the formation of the CME flux rope and its trajectory in the low corona are southward, but not nearly as southward as the filament-channel PIL ($-40^\circ$ latitude), due to the rapid expansion of the CME cross-section. A part of the CME flux rope intersects the ecliptic plane for almost the entirety of the eruption. Figure~\ref{fprop} shows the magnetic-field evolution in our system from the ``Earth viewpoint'' of the synthetic coronagraph data and observations corresponding to 2015 July 10. Figures~\ref{fprop}(a-d) are snapshots at exactly the same simulation times as the synthetic C3 panels in Figures~\ref{fwlsim}(e-h). From this vantage point, we see the complex, twisted field structure of the initial east-limb part of the eruption, how much of the structure is actually Earth-directed, and how the magnetic fields of the western leg of the CME open up (from east to west) toward the observer, giving rise to the flux-rope-like white-light signature on the west limb.

\begin{figure*}[!t]
	\centering{
	\includegraphics[width=0.95\textwidth]{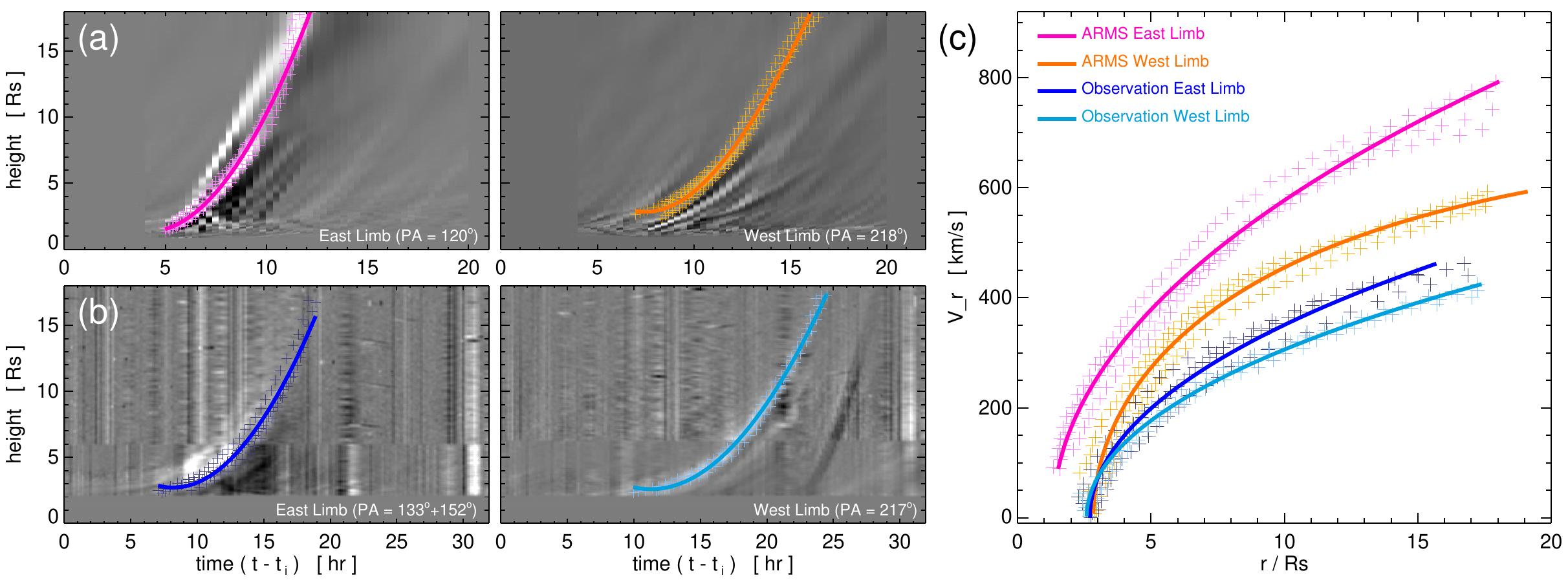}
	}
	\caption{Height--time J-maps and functional fits to the east- and west-limb parts of the global-scale CME eruption. (a): MHD simulation results from running-difference processing of the Figure~\ref{fwlsim} animation data. (b): Observational results from running-difference processing of the Figure~\ref{f2} animation data. (c): Analytic $v(r)$ profiles derived from the height--time fits.}
	\label{fht}
\end{figure*}

\begin{table*}[t]
\centering
\caption{Best-fit parameters for Equations~\ref{eq:ht} and \ref{eq:vr} for the simulated and observed height/time coronagraph data in Figure~\ref{fht}. The units of $t_0$ are hr, $r_0$ and $r_a$ are $R_\odot$, and $v_a$ and $v_{\rm fit}(20R_\odot)$ are km~s$^{-1}$.
	 \label{tab:htfit}}
\begin{tabular}{|l|ccccc|c|}
\tableline\tableline
    Height/time profile & $t_0$ & $r_0$ & $r_a$ & $v_a$ & $\chi^2$ & $v_{\rm fit}(20R_\odot)$ \\[0.02in]  
\tableline
ARMS East limb  & -175.1 &  1.32  & 190.4 & 2733.8 & 0.077 & 835.8 \\
ARMS West limb  & -178.5 &  2.84  & 13.4 &  707.4 & 0.115 & 601.0 \\
LASCO C2/C3 East limb  &  -24.2 &  2.71  & 110.1 &  1385.1 & 0.164 & 528.1 \\
LASCO C2/C3 West limb  &  -27.3 &  2.58  & 102.9 & 1160.1 & 0.051 & 457.8 \\
\tableline
\end{tabular}
\end{table*}

Quantitative comparisons between the simulated and observational data can be made with the eruption's height--time evolution and the resulting velocity profiles.
We fit both the simulation and observational height/time data with the \citet{Sheeley1999} function 

\begin{equation}
\label{eq:ht}
    h(t) = r_0 + 2 r_a \ln{\left[ \cosh{\left[ \frac{v_a(t+t_0)}{2r_a} \right]} \right]} \; .
\end{equation}

\noindent This height-time expression has four free parameters that describe the initial position ($r_0,t_0$), the asymptotic velocity ($v_a$), and the length scale ($r_a$) over which $v(r)$ reaches 80\% of $v_a$. The velocity profiles are then given as

\begin{equation}
\label{eq:vr}
    v(r) =  v_a \left( 1 - \exp{\left[ \frac{-(r-r_0)}{r_a} \right]} \right)^{1/2} \; .
\end{equation}

\noindent We used the IDL {\tt curvefit} function to minimize the weighted $\chi^2$ error between the parameterized $h_{\rm fit}(t)$ expression and the height--time data, 

\begin{equation}
    \chi^2 = \frac{1}{(N-4)}\sum_{i=0}^N w_i \big( h_{\rm fit}(t_i) - h(t_i) \big)^2 \; ,
\end{equation}

\noindent where the weights are simply $w_i = h^{-1}(t_i)/\max[h^{-1}(t_i)]$ and  $N-4$ is the number of data points in each profile minus the four free parameters. Figure~\ref{fht}(a) shows the the simulation height/time curves from the eastern limb at position angle (PA) of 120$^{\circ}$ and the western limb at PA=218$^{\circ}$ as the magenta and orange `+' symbols. The observational height--time curves for the LASCO data are shown in Figure~\ref{fht}(b) as the blue and cyan `+' symbols. We constructed the eastern limb LASCO height--time data as the sum of two PAs because of the location of the LASCO/C3 occulter arm: PAs 133$^{\circ}$ and 152$^{\circ}$ are on either side of the arm and capture the leading edge of the CME; see Figure~\ref{f2}(e--h). 
The best-fit parameters for each of the profiles are given in Table~\ref{tab:htfit}. 
The $v_{\rm fit}(r)$ curves are shown in Figure~\ref{fht}(c) in the same color scheme as the $h_{\rm fit}(t)$ profiles in \ref{fht}(a) and \ref{fht}(b). The last column of Table~\ref{tab:htfit} lists the $v_{\rm fit}$ values extrapolated to $r=20\,R_\odot$ for each profile.

The east-limb velocity profiles in both the simulation and observations are faster than their corresponding west-limb counterparts. 
This agreement suggests that, although the overall energy release in our model system is larger than that inferred from the 2015 July 9--10 observations, we can use the simulation results to infer the morphological evolution of the magnetic-field structure in the event. 
Specifically, we can interpret the evolution of the white-light structure in the coronagraph data as corresponding to the initial east-limb component of the CME generated during the impulsive phase of the eruption that shows a higher velocity, followed by the west-limb component that presents as the leg of the extended CME flux rope opening up towards the observer with a lower velocity.


\section{Discussion and Conclusions} \label{sec:disc}

The simulation results presented in this paper represent the first detailed calculation of a CME eruption (on 2015 July 9--10) from an extended, high-latitude PIL source region similar to the large filament-channel configurations examined by \citet{Pevtsov2012}. The studied PIL and the eruption extended all the way from the east to the west limb as seen from Earth. \citet{Anderson2005} have discussed the formation mechanisms of these huge filament channels. More recently, \citet{Patsourakos2020} have reviewed the magnetic structure of energized, pre-eruption states and concluded that there is likely a smooth distribution between more ``sheared-arcade-like'' and more ``flux-rope-like'' pre-eruptive configurations. The transition between the two may be an intrinsic aspect of CME initiation processes. A number of researchers have modeled the magnetic structure of these large, polar-crown filament channels (and their associated coronal cavities) using the flux-rope insertion method \citep{vanBallegooijen2004}. For example, \citet{Su2012}, \citet{Su2015}, and \citet{Jibben2016} have all determined best-fit model twist fluxes per unit length that range from zero to a few times $10^{10}$~Mx~cm$^{-1}$. Given the filament-channel lengths of 45$^{\circ}$--120$^{\circ}$ longitude, typical poloidal/twist flux estimates are on the order of 10$^{20\rm{-}21}$~Mx. Figure~\ref{fflside} shows that our model filament channel is clearly more of a ``sheared-arcade-like'' configuration, but the reconnection fluxes in both our simulation and the SDO observations result in poloidal/twist flux values entirely consistent with those earlier estimates. 

One of the first comprehensive Sun-to-heliosphere analyses of a particularly strong ``problem geomagnetic storm'' was the 1994 April 14 event discussed by \citet{McAllister1996}. The on-disk signatures for that extended filament-channel eruption included Yohkoh soft X-ray observations of a massive PEA that formed over the course of $\sim$10~hr to span $\sim$150$^{\circ}$ in longitude and 30$^{\circ}$--40$^{\circ}$ in latitude on the disk face. Ground-based coronagraph observations from K-Cor were able to resolve the helmet-streamer belt in the southern hemisphere overlying the extended PIL with indications of a streamer-blowout on the southeastern limb. \citet{McAllister1996} conclude, ``\emph{Thus, although there was no direct observation of a white-light CME on April~14, the existing data strongly support the conclusion that a CME did in fact take place.}'' 

\citet{Zhukov2007} have examined global-scale dimming signatures associated with X-class flares and their resulting halo CMEs. Their classification of a ``global'' eruption was based on EUV dimmings on the limb that were $\gtrsim 180^{\circ}$ in angular extent. There are a number of examples of sympathetic eruptions, where a CME triggers subsequent eruptions \citep[e.g.,][]{Torok2011, Schrijver2013}, and this scenario may explain global EUV dimming observations, at least in some cases. In the event modeled here, both the observations and simulation present EUV signatures suggesting that an eruption occurred across a comparably sized source region. However, there was no obvious off-limb dimming signature, nor were the SDO/AIA 211~{\AA} dimmings especially dramatic, despite the event being a clearly ``global'' eruption.

Forward-modeling of synthetic white-light images using the simulation's mass-density data enabled us to interpret the coronagraph observations during this event in a particularly illuminating way. Specifically, the seemingly disparate CME events from opposite limbs, listed as two separate CMEs occurring over 7 hours apart in the LASCO CME catalog, have been shown here to be consistent with a single, gradual eruption from a high-latitude PIL. The 2015 July 9--10 large-scale filament eruption can be considered an asymmetric eruption of the type discussed by, e.g., \citet{Tripathi2006} and \citet{LiuR2009}. In both the observations and our simulation, the east-limb magnetic fields erupt first, and the eruption thereafter proceeds to sweep from east to west along the PIL. 

\cite{McCauley2015} created a catalog of SDO prominence and filament eruptions and examined the statistics for events organized by various properties and source-region types (i.e., active region, intermediate, quiescent, and polar crown). For example, symmetric filament eruptions are slightly more common (48\%) than asymmetric eruptions (38\%) over the whole data set and in each source-region type, with the exception of polar-crown filaments whose eruptions were 39\% symmetric and 45\% asymmetric. \citet{McCauley2015} suggested that the longer filaments may provide more opportunity for localized destabilization, which then propagates along the energized filament channel---in excellent qualitative agreement with our simulation results.

Polar-crown and quiet-Sun erupting filaments are known to deflect away from coronal holes and toward the equator and/or the streamer belt and HCS \citep[e.g.,][]{Kilpua2009AnGeo, Panasenco2013}. In the \citet{McCauley2015} survey, approximately 25\% of filament eruptions propagate non-radially or exhibit an even more extreme deflection (e.g., erupting sideways). \citet{Gopalswamy2015} showed that deflections up to $\sim$30$^{\circ}$ in the LASCO coronagraph field of view are fairly common for polar-crown filament CMEs. The 2015 July 9--10 CME does not appear to have a significant deflection during the eruption in either observations or our simulation. In part, this is because the filament channel is comfortably under a high-latitude excursion of the helmet-streamer belt. Thus, our eruption is more of a classic streamer-blowout CME \citep{Lynch2016b,Vourlidas2018} that happens to originate from the helmet-streamer belt at $-40^{\circ}$ latitude. 

However, despite the high-latitude source region, the rapidly expanding CME eruption becomes large enough to intersect the equatorial plane. This includes the top of the coherent, magnetic flux-rope ejecta which is easily seen in the simulation magnetic-field evolution (Figures~\ref{fflside} and \ref{fprop}), but neither the observed or synthetic coronagraph imagery show an obvious component of the CME in the ecliptic headed toward Earth.  While the initial east-limb white-light structure of our model eruption is larger and more equatorial than the observations, the west-limb part of the simulation CME is in much better agreement with the apparent propagation direction (cf.\ Figures~\ref{f2} and \ref{fwlsim}). 

The intersection with the ecliptic plane is important because this high-latitude filament-eruption CME impacted Earth. The ICME was observed by the Wind and ACE spacecraft beginning on 2015 July 13, and the in-situ magnetic-field and plasma measurements exhibit various properties compatible with typical magnetic-cloud/flux-rope signatures. Additionally, this particular ICME had a sustained period of southward $B_z$, which caused a moderate geomagnetic disturbance (peak minimum Dst of $-61$~nT). Both the impact and the geoeffectiveness of this event may have been unexpected, given the seemingly disparate eruptions in the LASCO data that were both slow and apparently directed more southward than toward Earth.

The original definition of a ``stealth CME'' is an event that lacks clear, on-disk eruption signatures but can be visible in coronagraph observations, especially away from the Sun--Earth line. The opposite scenario also happens: There can be eruption signatures on the disk but no readily identifiable or significant CME counterpart in coronagraph data. The 2015 July 9--10 event and our simulation fall into a third, much more general category: ambiguous on-disk eruption signatures with ambiguous coronagraph signatures. Here, the ``stealthiness'' is not missing observational signatures; rather, it represents the uncertainty in assessing the likelihood of an Earth impact and/or its geoeffectiveness (i.e., what makes a problem geomagnetic storm ``problematic''). In a future paper, we intend to analyze the interplanetary propagation of this high-latitude filament-channel eruption, examine its in-situ plasma, field, and composition characteristics, and model its Sun-to-Earth evolution.


\acknowledgments

The authors are pleased to acknowledge the productive and fruitful collaboration of the International Space Science Institute (ISSI) Team \#415 ``Understanding the Origins of Problem Geomagnetic Storms'' led by N.~Nitta and T.~Mulligan (\url{https://www.issibern.ch/teams/geomagstorm/}), from which this work originated.
B.J.L.\ and M.D.K.\ acknowledge support from NSF AGS-1622495, NASA NNX17AI28G, and NASA 80NSSC19K0088.
E.P.\ acknowledges support from the Oskar {\"O}flund Foundation and the NASA Living With a Star Jack Eddy Postdoctoral Fellowship Program, administered by UCAR's Cooperative Programs for the Advancement of Earth System Science (CPAESS) under award NNX16AK22G.
C.R.D.\ acknowledges NASA's H-ISFM and H-LWS programs.
J.T.D. acknowledges support from the NASA Postdoctoral Program at the NASA Goddard Space Flight Center, administered by Universities Space Research Association under contract with NASA.
J.P.\ and E.K.J.K.\ acknowledge the European Research Council (ERC) under the European Union's Horizon 2020 Research and Innovation Programme Project SolMAG (grant agreement no. 724391) and the Finnish Centre of Excellence in Research of Sustainable Space (Academy of Finland grant no. 312390).

The NASA Solar Dynamics Observatory data are provided by the SDO/AIA and SDO/HMI instrument teams (\url{https://sdo.gsfc.nasa.gov/}).
The NSO/GONG data are provided by the National Solar Observatory, operated by AURA under a cooperative agreement with NSF, and with additional financial support from NOAA, NASA, and USAF (\url{https://gong.nso.edu/}).
The LASCO CME catalog is generated and maintained at the CDAW Data Center by NASA and The Catholic University of America in cooperation with the Naval Research Laboratory (\url{https://cdaw.gsfc.nasa.gov/CME_list/}).
The SOHO/LASCO data used here are produced by a consortium of the Naval Research Laboratory (USA), Max-Planck-Institut for Sonnensystemforschung (Germany), Laboratorie d'Astrophysique Marseille (France), and the University of Birmingham (UK). SOHO is a project of international cooperation between ESA and NASA (\url{http://lasco-www.nrl.navy.mil/}). 
The K-Cor coronagraph data are provided by the Mauna Loa Solar Observatory, operated by the High Altitude Observatory, as part of the National Center for Atmospheric Research (NCAR). NCAR is supported by the NSF (\url{https://www2.hao.ucar.edu/mlso/mlso-home-page}).

\facilities{COSMO (K-Cor); NSO (GONG); SDO (AIA, HMI); SOHO (LASCO)}

\software{ARMS \citep{DeVore2008}; SolarSoft \citep{Freeland1998}; JHelioviewer \citep{Mueller2017}; SunPy \citep{Sunpy2020}}

\clearpage


\appendix

\section{STITCH Energization Parameters}
\label{appendix:driving}

The STITCH formalism represents the introduction of horizontal magnetic flux at the lower boundary, allowing for the accumulation of a non-potential shear/twist component of the magnetic field concentrated above our large-scale polarity inversion line.
{The details and justification of the physical basis for the statistically averaged helicity condensation model and its implementation for modeling the long-term, non-potential evolution of the global solar corona can be found in the appendix of \citet{Mackay2014}.}

In ARMS, the STITCH contribution is implemented as a source term in the induction equation, contributing new horizontal components of the magnetic field, $\partial \boldsymbol{B}_S/\partial t$, given by Equation~\ref{eq:stitch}. The $B_\theta$, $B_\phi$ components of the magnetic field are defined at their respective cell faces. The STITCH updates are applied to the bottom-most computational grid cells in the domain and are then transported further upward into the domain by convection. The radial component, $B_r$, on the inner radial boundary (half a grid cell below the $B_\theta$, $B_\phi$ components) remains unchanged by the STITCH energization.
The STITCH contributions are calculated from the curl of a set of functions $\zeta^{(i)} B_r(\theta, \phi) \boldsymbol{\hat{r}}$. The form of $\zeta^{(i)}$ is given in Equation~\ref{eq:zeta} as the product of spatial and temporal envelope functions $f_\theta(\theta)$, $f_\phi(\phi)$, $f_t(t)$ applied to regions of positive or negative magnetic polarity via $f_B(B_r)$.
The $\theta$ and $\phi$ envelope functions are given by

\begin{equation}
    f_\theta(\theta) = \onehalf - \onehalf \cos{\left[ 2\pi k_\theta \frac{(\theta-\theta_c)}{(\theta_r - \theta_l)} \right]} \;,
\end{equation} 

\begin{equation}
f_\phi(\phi) = 
    \begin{cases}
        \sin{\left[ 2 \pi k_\phi \frac{(\phi-\phi_c)}{(\phi_r - \phi_l)}  \right]} & \text{for } i=1,2 \\
        \onehalf - \onehalf \cos{\left[ 2\pi k_\phi \frac{(\phi-\phi_c)}{(\phi_r - \phi_l)} \right]} & \text{for } i \ge 3
    \end{cases}
\end{equation}

\noindent and the temporal dependence is given by

\begin{equation}
    f_t(t) = \onehalf - \onehalf \cos{\left[ 2\pi k_t \frac{(t-t_c)}{(t_r-t_l)} \right]} \; .
\end{equation}

\noindent {The $l$, $r$, and $c$ subscripts just refer to the \textit{left}, \textit{right}, and \textit{centering} values that define the spatial or temporal range of the $\theta$, $\phi$, and $t$ envelope functions.} The masking of the $B_r$ polarities is specified by

\begin{equation}
f_B(B_r) = 
    \begin{cases}
        1 & \text{for } B_r/|B_r| = P_B \\
        0 & \text{elsewhere}
    \end{cases} \; .
\end{equation}

\noindent We use $N = 8$ separate $\zeta^{(i)}(B_r, \theta,\phi,t)$ patterns that are added together to generate the final helicity-injection distribution. Each of the parameter sets for the spatial and temporal functions are given in Table~\ref{tab:stitch}. As described in Sections~\ref{subsec:flows} and \ref{subsec:filament}, we performed two phases of helicity injection and magnetic-field energization with this pattern: the first for $100 {\rm \; hr} \le t \le 130 {\rm \; hr}$, and the second for $160 {\rm \; hr} \le t \le 180 {\rm \; hr}$. The parameters used in the second phase, `Activation and Eruption,' are listed in parentheses where they differ from the parameters of the first phase, `Filament-channel Energization.'

\begin{figure*}
	\centerline{ \includegraphics[width=0.475\textwidth]{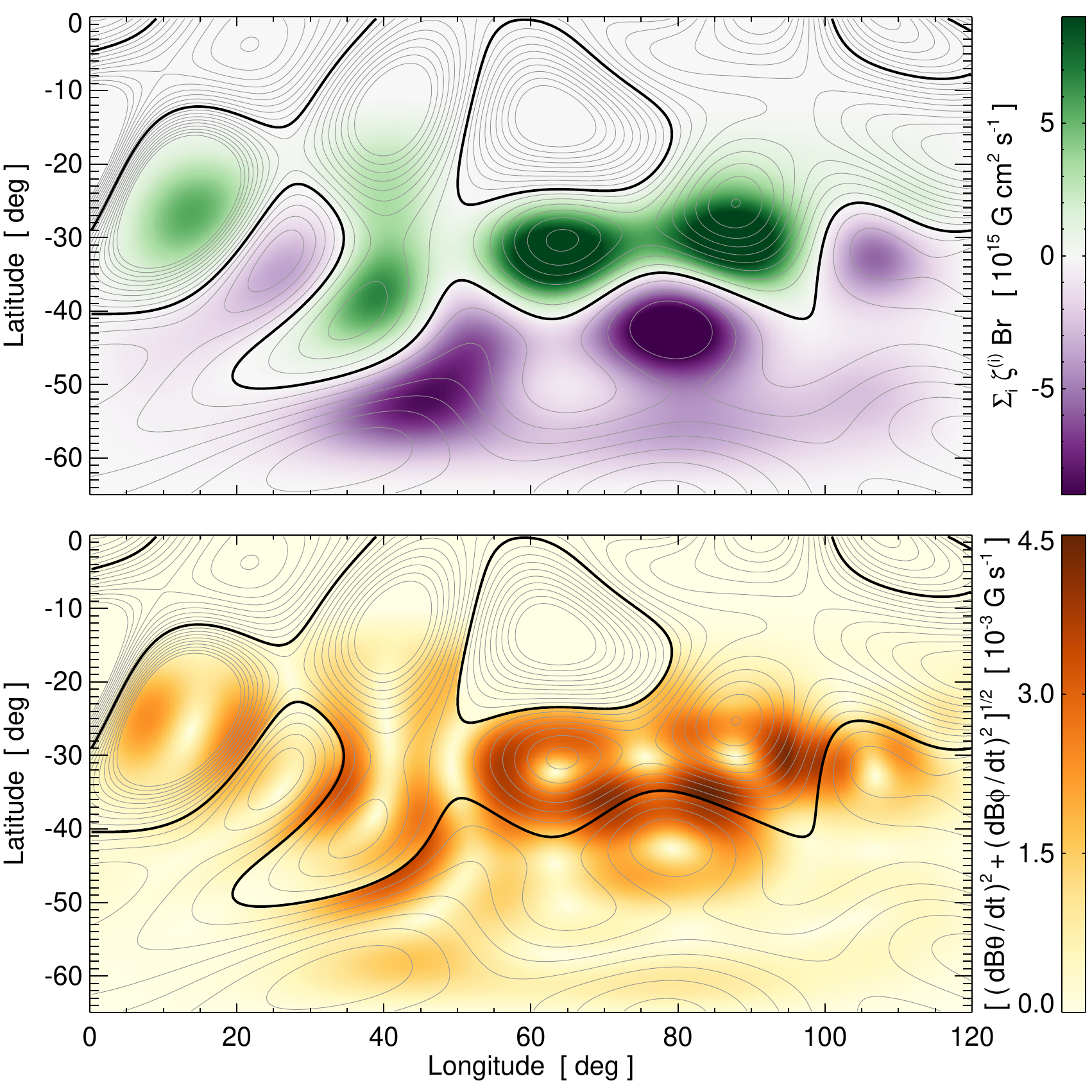}}
	\caption{The top panel shows the summation of the eight different $\zeta^{(i)} B_r$ contributions used to construct the STITCH energization boundary conditions. The bottom panel shows the magnitude of the STITCH horizontal field changes $[ (\partial B_\theta/\partial t)^2 + (\partial B_\phi/\partial t)^2 ]^{1/2}$.}
	\label{fzetas}
\end{figure*}

Figure~\ref{fzetas}(a) shows the resulting $\sum \zeta^{(i)} B_r$ distribution at its maximum during the Filament-channel Energization phase, when $f_t( 115~{\rm hr} )=1.0$. 
The magnitudes of the four STITCH patterns acting on the positive-polarity regions ($i=1, 3, 5, 7$) are shown in green, while the negative-polarity patterns ($i=2, 4, 6, 8$) are shown in purple. 
{Figure~\ref{fzetas}(b) shows the magnitude of the right-hand side of  Equation~\ref{eq:stitch} in units of $10^{-3}$~G~s$^{-1}$.} 
The magnitude pattern for $50^{\circ} \le \phi \le 120^{\circ}$ shows a series of rings 
that reflect the $\zeta^{(i)} B_r$ component patterns defined above. 
The individual $\partial B_\theta/\partial t$ and $\partial B_\phi/\partial t$ components that make up the Figure~\ref{fzetas}(b) helicity-injection magnitude are shown in Figure~\ref{fstitch}. These components, when plotted separately, illustrate the cumulative effect of multiple helicity-injection patterns creating a coherent, sheared-field filament channel above the complex, high-latitude PIL.

\begin{table}
\centering
\caption{Parameters for the spatial and temporal dependence of $\zeta^{(i)}(\theta,\phi,t)$.
	 \label{tab:stitch}}
\begin{tabular}{|r|rr|rrrr|rrrr|rrrr|c|}
\tableline\tableline
     $\;$ & $\;$ & $\;$ 
     & \multicolumn{4}{c|}{$f_{\theta}(\theta)$} 
     & \multicolumn{4}{c|}{$f_{\phi}(\phi)$}
     & \multicolumn{4}{c|}{$f_{t}(t)$}
     & $f_{B}(B_r)$ \\
   $i$ & \multicolumn{2}{c|}{$K_0$\tablenotemark{a}} &
   $k_\theta$   & $\theta_l$ [$^\circ$]    & $\theta_r$ [$^\circ$]    & $\theta_c$ [$^\circ$]  & 
   $k_\phi$  & $\phi_l$ [$^\circ$]       & $\phi_r$ [$^\circ$]      & $\phi_c$ [$^\circ$]  & 
   $k_t$        & $t_l$ [hr]         & $t_r$ [hr]        & $t_c$ [hr]  &
   $P_B$ \\
\tableline
1  & $ 1.0$ &  $(0.50)$ &  1.0  & $-55$ & $-10$ & $-10$    & 0.5 &  0 & 120 &  0  & 1.0 & 100 (160) & 130 (180) & 100 (160) & $+1$ \\
2  & $-1.5$ & $(-0.75)$ &  1.0  & $-65$ & $-20$ & $-20$    & 0.5 &  0 & 120 &  0  & 1.0 & 100 (160) & 130 (180) & 100 (160) & $-1$ \\
3  &  $1.0$ &  $(0.50)$ &  1.0  & $-55$ & $-20$ & $-20$    & 2.0 & 20 & 120 & 20  & 1.0 & 100 (160) & 130 (180) & 100 (160) & $+1$ \\
4  & $-1.0$ & $(-0.50)$ &  1.0  & $-65$ & $-20$ & $-20$    & 2.0 & 20 & 120 & 20  & 1.0 & 100 (160) & 130 (180) & 100 (160) & $-1$ \\
5  &  $1.5$ &  $(0.75)$ &  1.0  & $-45$ & $-20$ & $-20$    & 1.0 & 95 & 120 & 95  & 1.0 & 100 (160) & 130 (180) & 100 (160) & $+1$ \\
6  & $-1.5$ & $(-0.75)$ &  1.0  & $-45$ & $-20$ & $-20$    & 1.0 & 95 & 120 & 95  & 1.0 & 100 (160) & 130 (180) & 100 (160) & $-1$ \\
7  &  $1.5$ &  $(0.75)$ &  1.0  & $-51$ & $-23$ & $-23$    & 1.0 & 20 & 120 & 20  & 1.0 & 100 (160) & 130 (180) & 100 (160) & $+1$ \\
8  & $-1.5$ & $(-0.75)$ &  1.0  & $-65$ & $-25$ & $-25$    & 1.0 & 20 & 120 & 20  & 1.0 & 100 (160) & 130 (180) & 100 (160) & $-1$ \\
\tableline
\end{tabular}
\tablenotetext{a}{Units of the scalar coefficient are $10^{15}$~cm$^{2}$~s$^{-1}$.}
\end{table}

\section{AIA Flare Ribbon/Post-eruption Arcade Pixel Masks}
\label{appendix:obsflare}

The identification and tracking of flare ribbons is often done with 1600~{\AA} UV emission data \citep[e.g.,][]{Qiu2009}. \citet{Kazachenko2017} developed an empirical thresholding technique to compensate for CCD saturation and pixel blooming, in order to smoothly accumulate the total area swept out by the flare-ribbon motions. 
As discussed in Section~\ref{subsec:disk_obs}, however, the 2015 July 9--10 filament eruption does not generate these traditional flare-ribbon emission signatures. 
On the other hand, Figure~\ref{f1} and its animation show EUV signatures that are suggestive of ribbon-like evolution, but are more clearly indicative of the post-eruption arcade (PEA) brightening and the refilling of eruption-related EUV dimmings. We note that the area underneath the PEA can be used to estimate the reconnection flux \citep[see, e.g.,][]{Gopalswamy2017}, but this technique is typically used at a single time corresponding to the maximum area of the PEA emission. 

\begin{figure*}
	\centerline{ \includegraphics[width=0.95\textwidth]{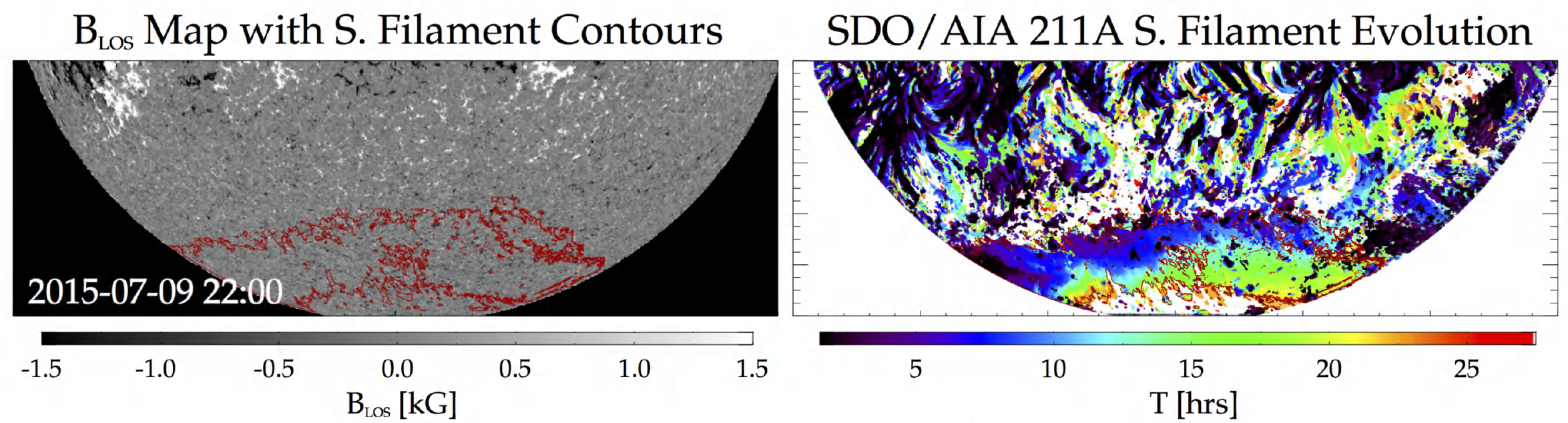}}
	\caption{The left panel shows the post-eruption arcade area (red contour) used to estimate the negative polarity reconnection flux $\Phi^{211}_{\rm rxn}$ from the HMI $B_{\rm LOS}$ data. The right panel shows the cumulative ribbon area, with each pixel colored according to the first time the reconnection threshold was exceeded in the running-difference AIA 211~{\AA} image.}
	\label{fobsmask}
\end{figure*}

{We used a 48-hour sequence of AIA images in the 211~{\AA} channel, from 2015 July 9 at 16:00~UT until 2015 July 11 at 15:30~UT with a 30-minute temporal cadence and a 0.6'' pixel size, to track the evolution of the PEA boundaries.} Since the evolution was not well seen in individual images, we used difference images 30 minutes apart to track the boundary motions. 
We smoothed each difference image with a $10 \times 10$ pixel boxcar average and identified pixels where the intensity was greater than three exposure-normalized data counts in the difference image to construct the cumulative pixel mask array $M_{i,j}$. 
We then rotated the derived filament mask sequence using {\tt diff\_rot.pro} in SolarSoft to the first image when the filament started rising (2015 July 9 at 23:00 UT) to compensate for the solar rotation. 

Figure~\ref{fobsmask} shows the evolution of the identified southern portion of the PEA. Panel (a) shows the HMI line-of-sight (LOS) field $B_{\rm LOS}$ with the cumulative ribbon mask area denoted by the red contours. Panel (b) shows the time at which each individual pixel met the running-difference intensity threshold.
We were not able to track the flare arcade area north of the PIL unambiguously, due to the presence of bright emission from several large active regions.
We calculated the evolution of the total reconnection flux associated with the southern half of the filament channel from the cumulative ribbon mask $M_{i,j}^{211}$ corresponding to the negative polarity region $(B_r \le 0)$ as  

\begin{equation}
\Phi_{\rm rxn}^{211} = \int |B_r^{(-)}| \; dA^{211} = \sum_{i,j}  \frac{ | B_{\rm LOS}^{(-)} | }{\cos^2{\theta}}  \; M_{i,j}^{211} \; dA_{i,j} \; .
\end{equation}

\noindent In the magnetic flux calculation, we only considered magnetic fields above the noise level $B_\mathrm{LOS}=15$~G. The $\cos^2{\theta}$ factor comes from de-projecting the LOS magnetic fields onto the radial direction $B_r^{(-)} = B^{(-)}_\mathrm{LOS}/\cos{\theta}$ and taking into account foreshortening of the surface area in each pixel $(dA^{211} = dA_{i,j}/\cos{\theta})$.  
Figure~\ref{fobsmask}(b) shows the classic time-integrated two-ribbon-flare evolution signatures \citep{Qiu2009, Aulanier2012, Kazachenko2017}, but for just the southern ribbon. This reconnection-flux area quickly extends from east to west during $5 {\rm \; hr}\; \lesssim t \lesssim 10$~hr, corresponding to rapid growth parallel to the global PIL. Afterwards, the southern half of the PEA area grows more slowly during $10 {\rm \; hr} \lesssim t\lesssim 25$~hr, perpendicular to the global PIL. 
Our reconnection flux estimate during the filament eruption is shown in Figure~\ref{feuv2}(b) as the black line with square symbols. 
The reconnection rate is defined as $d\Phi_{\rm rxn}^{211}/dt$, calculated using the standard central-difference formula. Our reconnection rate estimate is shown in Figure~\ref{feuv2}(b) as the purple line.

\bibliographystyle{aasjournal}
\bibliography{ep}

\end{document}